\documentclass[aps,prd,amsmath,amssymb,eqsecnum,reprint,showpacs,groupedaddress]{revtex4-1}

\usepackage{color}
\usepackage{graphicx}

\newcommand{\bmath}[1]{\mbox{{\boldmath{{$#1$}}}}}

\begin{document}

\author{Ben L. Shepherd}
\affiliation{Consortium for Fundamental Physics,
School of Mathematics and Statistics, The University of Sheffield,
Hicks Building, Hounsfield Road, Sheffield.~S3 7RH United Kingdom}

\author{Elizabeth Winstanley}
\email{E.Winstanley@sheffield.ac.uk}
\affiliation{Consortium for Fundamental Physics,
School of Mathematics and Statistics, The University of Sheffield,
Hicks Building, Hounsfield Road, Sheffield.~S3 7RH United Kingdom}

\title{Dyons and dyonic black holes in ${\mathfrak {su}}(N)$
Einstein-Yang-Mills\\ theory in anti-de Sitter space-time}

\begin{abstract}
We present new spherically symmetric, dyonic soliton and black hole solutions of the ${\mathfrak {su}}(N)$ Einstein-Yang-Mills equations in four-dimensional asymptotically anti-de Sitter space-time.
The gauge field has nontrivial electric and magnetic components and is described by $N-1$ magnetic gauge field functions and $N-1$ electric gauge field functions.
We explore the phase space of solutions in detail for ${\mathfrak {su}}(2)$ and ${\mathfrak {su}}(3)$ gauge groups.
Combinations of the electric gauge field functions are monotonic and have no zeros; in general the magnetic gauge field functions may have zeros.
The phase space of solutions is extremely rich, and we find solutions in which the magnetic gauge field functions have more than fifty zeros.
Of particular interest are solutions for which the magnetic gauge field functions have no zeros, which exist when the negative cosmological constant has sufficiently large magnitude.
We conjecture that at least some of these nodeless solutions may be stable under linear, spherically symmetric, perturbations.
\end{abstract}


\date{\today }

\maketitle

\section{Introduction}
\label{sec:intro}

The study of soliton and black hole solutions of the Einstein-Yang-Mills (EYM) equations has been an active subject for some twenty-five years,
triggered by the discovery of regular, static, spherically symmetric, solitons \cite{Bartnik:1988am} and ``coloured'' black holes \cite{Bizon:1990sr} in ${\mathfrak {su}}(2)$ EYM in four-dimensional asymptotically flat space-time. In these configurations, the nonabelian gauge field is purely magnetic \cite{Ershov:1991nv} and described by a single function $\omega _{1}(r)$ of the radial coordinate $r$.
Numerical investigations \cite{Bartnik:1988am, Bizon:1990sr} show that $\omega _{1}(r)$ must have at least one zero, and this is has also been proven analytically
\cite{Breitenlohner:1993es}.
Both the particle-like and black hole solutions arise at discrete points in the phase space of parameters, and can be indexed by $n$, the number of zeros of $\omega _{1}(r)$.
The ${\mathfrak {su}}(2)$ solitons and black holes have no magnetic charge at infinity.
The black holes, in particular, are indistinguishable from a Schwarzschild black hole at infinity, and therefore present counter-examples to the ``no-hair'' conjecture \cite{Ruffini:1971bza}.
However, both the black hole and soliton solutions are unstable under linear, spherically symmetric, perturbations of the gauge field and metric \cite{Straumann:1989tf}.
As a consequence of this, Bizon postulated a modified ``no-hair'' conjecture \cite{Bizon:1994dh}, heuristically stated as ``a stable black hole is uniquely characterized by global charges''.  The ``coloured'' black holes do not contradict this conjecture due to their instability.

Notwithstanding the instability of the ${\mathfrak {su}}(2)$ EYM solitons and black holes in asymptotically flat space-time, their discovery sparked extensive research into soliton and black hole solutions of more general matter models involving nonabelian gauge fields (see \cite{Volkov:1998cc} for a review).
In four-dimensional asymptotically flat space-time, enlarging the gauge group to ${\mathfrak {su}}(N)$, a richer phase space arises  \cite{Galtsov:1991au}, but solutions still occur at discrete points in the phase space.
Furthermore, there is a very general result that all purely magnetic, asymptotically flat, soliton or black hole solutions of the ${\mathfrak {su}}(N)$ EYM equations in four-dimensional space-time are unstable \cite{Brodbeck:1994np}.

The situation is very different if one considers EYM with a negative cosmological constant $\Lambda <0$, so that the space-time is asymptotically anti-de Sitter (adS) rather than asymptotically flat.
Four-dimensional, purely magnetic, black hole \cite{Winstanley:1998sn, Bjoraker:1999yd} and soliton \cite{Bjoraker:1999yd, Breitenlohner:2003qj} solutions of the ${\mathfrak {su}}(2)$ EYM equations in adS are found in continuous regions of parameter space rather than at discrete points as in the asymptotically flat case.
Furthermore, if $\left| \Lambda \right|$ is sufficiently large, there are solutions for which the single magnetic gauge field function $\omega _{1}(r)$ has no zeros.
These nodeless solutions have no analogue in asymptotically flat space-time and at least some of them are stable under linear, spherically symmetric perturbations \cite{Winstanley:1998sn, Bjoraker:1999yd}.
The proof of stability can be extended to general linear perturbations of the metric and gauge field providing $\left| \Lambda \right| $ is sufficiently large
\cite{Sarbach:2001mc}.

For the larger ${\mathfrak {su}}(N)$ gauge group, purely magnetic soliton and black hole solutions of EYM in adS exist in continuous regions of the parameter space \cite{Baxter:2007at}. In this case the purely magnetic gauge field is described by $N-1$ functions $\omega _{j}(r)$, $j=1,\ldots N-1$, of the radial coordinate $r$.
There exist both soliton and black hole solutions for which all the $\omega _{j}(r)$ have no zeros provided $\left| \Lambda \right| $ is sufficiently large \cite{Baxter:2008pi}.
These solutions are of particular interest because it can be proven that at least some of them are stable under linear, spherically symmetric, perturbations of the metric and gauge field \cite{Baxter:2015gfa}.
Although these stable black holes are coupled to potentially unlimited amounts of gauge field hair, they can nonetheless be characterized by global charges defined far from the black hole \cite{Shepherd:2012sz}, in accordance with the modified ``no-hair'' conjecture \cite{Bizon:1994dh} (see also the reviews
\cite{Winstanley:2008ac}).

Returning to ${\mathfrak {su}}(2)$ EYM in adS, the space of solutions revealed another surprise \cite{Bjoraker:1999yd}.  In asymptotically flat space-time, all nontrivial (that is, not corresponding to embedded Schwarzschild or abelian Reissner-Nordstr\"om configurations) ${\mathfrak {su}}(2)$
solitons and black holes must have gauge fields with vanishing electric parts \cite{Ershov:1991nv}.  This is not the case in asymptotically adS space-time,
with dyonic solitons and black holes existing, for which the gauge field has nontrivial electric and magnetic components \cite{Bjoraker:1999yd}.
For ${\mathfrak {su}}(2)$ gauge group, the electric part is described by a single function $h_{1}(r)$ and the magnetic part by a single function $\omega _{1}(r)$, where $r$ is the radial coordinate.
The electric gauge field function $h_{1}(r)$ is monotonic and has no zeros;
there exist solitons and black holes for which the magnetic gauge field function $\omega _{1}(r)$ also has no zeros \cite{Bjoraker:1999yd, Nolan:2012ax}.
Although these solutions were found numerically fifteen years ago \cite{Bjoraker:1999yd}, it was only recently that it was shown that at least some of the solutions for which $\omega _{1}(r)$ is nodeless are stable under linear, spherically symmetric perturbations \cite{Nolan:2015vca}.

In this paper we explore the phase space of dyonic EYM solitons and black holes in adS when the gauge group is enlarged to ${\mathfrak {su}}(N)$.
In the purely magnetic case, with the larger gauge group the phase space has a very rich structure \cite{Baxter:2007at} and we anticipate that the same will be true for dyonic solutions.
Dyonic solutions of ${\mathfrak {su}}(N)$ EYM will be described by $N-1$ electric gauge field functions $h_{j}(r)$, $j=1,\ldots , N-1$, and $N-1$
magnetic gauge field functions $\omega _{j}(r)$, $j=1,\ldots , N-1$.
Recently the existence of soliton and black hole solutions (when $\left| \Lambda \right| $ is sufficiently large) for which all the  $\omega _{j}$ functions have no zeros was proven \cite{Baxter:2015tda}.
These nodeless solutions are of particular interest because, in the light of results in the purely magnetic sector \cite{Baxter:2015gfa} and for ${\mathfrak {su}}(2)$
dyonic solutions \cite{Nolan:2015vca}, it is anticipated that at least some of them will be stable under linear, spherically symmetric perturbations.
In this paper we present the first numerical dyonic solutions of the ${\mathfrak {su}}(N)$ EYM equations in adS, focusing not just on the nodeless solutions
whose existence is proven in \cite{Baxter:2015tda}, but also more generally on the rich features of the phase space of solutions.

The outline of this paper is as follows.
In Sec.~\ref{sec:equations} we introduce the ${\mathfrak {su}}(N)$ EYM model and field equations, and describe our static, spherically symmetric ansatz for the metric and gauge field potential.  We also discuss the boundary conditions at the origin (for regular solitons), event horizon (for black holes) and at infinity.
Solutions of the field equations are discussed in detail in Sec.~\ref{sec:solutions}.  We describe our numerical method, and explore the phase space of dyonic soliton and black hole solutions with ${\mathfrak {su}}(2)$ and ${\mathfrak {su}}(3)$ gauge group.
Our conclusions are presented in Sec.~\ref{sec:conc}.

\section{${\mathfrak {su}}(N)$ Einstein-Yang-Mills theory in anti-de Sitter space-time}
\label{sec:equations}

In this section we describe our gauge field and metric ansatz, the field
equations and the boundary conditions
for static, spherically symmetric dyon and dyonic black hole solutions
of ${\mathfrak {su}}(N)$ EYM in adS.

\subsection{Ansatz and field equations}
\label{sec:ansatzeqns}

\subsubsection{Metric and gauge potential ansatz}
\label{sec:ansatz}

We consider static, spherically symmetric, four-dimensional solitons
and black holes with metric
\begin{equation}
ds^{2} = -\sigma ^{2} \mu \, dt^{2} + \mu ^{-1} \, dr^{2}
+ r^{2} \left( d\theta ^{2} + \sin ^{2} \theta \, d\phi^{2} \right) ,
\label{eq:metric}
\end{equation}
where the metric functions $\mu $ and $\sigma $ depend only on the radial coordinate
$r$.
We write the metric function $\mu (r)$ in the form
\begin{equation}
\mu (r) = 1 -\frac{2m(r)}{r} - \frac{\Lambda r^{2}}{3},
\label{eq:mu}
\end{equation}
where $\Lambda <0$ is the cosmological constant.
In (\ref{eq:metric}) and throughout this paper, the metric has signature
$\left( -, +, +, + \right) $ and we use units in which $4\pi G = 1 = c$.

For the static, spherically symmetric ${\mathfrak {su}}(N)$ gauge field
potential $A$,
we use the ansatz \cite{Kuenzle:1991wa}
\begin{eqnarray}
A & = & {\mathcal {A}} \, dt + {\mathcal {B}} \, dr
+\frac {1}{2} \left( C-C^{H} \right) d\theta
\nonumber \\ & &
-\frac {i}{2} \left[ \left( C+C^{H} \right) \sin \theta
+D\cos \theta \right] d\phi ,
\label{eq:gaugepot}
\end{eqnarray}
where we have set the gauge coupling $g=1$.
In (\ref{eq:gaugepot}), ${\mathcal {A}}$, ${\mathcal {B}}$, $C$ and $D$
are $N \times N $ matrices which depend only on the radial coordinate $r$.
The matrices ${\mathcal {A}}$ and ${\mathcal {B}}$ are purely imaginary, diagonal
and traceless.
The matrix $C$ (with Hermitian conjugate $C^{H}$) is upper triangular, with the only nonzero
entries immediately above the diagonal, which are written as:
\begin{equation}
C_{j,j+1} = \omega _{j}(r) e^{i\gamma _{j}(r)} ,
\qquad
j= 1, \ldots , N-1,
\label{eq:Cj}
\end{equation}
where $\omega _{j}(r)$ and $\gamma _{j}(r)$ are real functions.
The matrix $D$ is constant, diagonal and traceless, and given by
\begin{equation}
D = {\mathrm {Diag}} \left( N-1, N-3,..., 3-N, 1-N \right) .
\label{eq:D}
\end{equation}
The ansatz (\ref{eq:gaugepot}) has some residual gauge freedom which can be
used to set ${\mathcal {B}}\equiv 0$ \cite{Kuenzle:1991wa}.
It should be emphasized that the choice of ansatz (\ref{eq:gaugepot}) is not unique
\cite{Bartnik:1997my}.

For purely magnetic solutions \cite{Baxter:2007at,Baxter:2008pi}
one makes the choice ${\mathcal {A}}\equiv 0$.
However, in this paper we are interested in solutions for which the gauge field
has nontrivial electric and magnetic parts.
Therefore ${\mathcal {A}}$ will not vanish.
We write ${\mathcal {A}}$ in terms of matrices $H_{\ell}$, which are the diagonal
generators of the Cartan subalgebra of ${\mathfrak {su}}(N)$.
We define the $H_{\ell}$ in a similar way to Ref.~\cite{Brandt:1980em}, but using
a slightly different normalization.
The nonzero entries of the matrix $H_{\ell}$ are:
\begin{equation}
\left[ H_{\ell} \right] _{j,k} = - \frac {i}{{\sqrt {2\ell \left( \ell +1 \right) }}} \left[
\sum _{p=1}^{\ell} \left( \delta _{j,p} \delta _{k,p} \right)
- \ell \delta _{j, \ell+1} \delta _{k,\ell+1} \right] ,
\label{eq:Hdef}
\end{equation}
where $\delta _{j,k}$ is the usual Kronecker delta.
For ${\mathfrak {su}}(2)$ EYM, there is a single generator of the Cartan subalgebra:
\begin{equation}
H_{1} =
- \frac {i}{2}
\left(
\begin{array}{cc}
1 & 0 \\
0 & -1
\end{array}
\right) ,
\label{eq:su2H1}
\end{equation}
while for ${\mathfrak {su}}(3)$ EYM, there are two $H$ matrices:
\begin{equation}
H_{1} =
-\frac {i}{2}
\left(
\begin{array}{ccc}
1 & 0 & 0 \\
0 & -1 & 0 \\
0 & 0 & 0
\end{array}
\right) ,
\qquad
H_{2} =
-\frac {i}{2{\sqrt {3}}}
\left(
\begin{array}{ccc}
1 & 0 & 0 \\
0 & 1 & 0 \\
0 & 0 & -2
\end{array}
\right) .
\label{eq:su3H}
\end{equation}
For general $N$, there are $N-1$ matrices $H_{\ell}$ (\ref{eq:Hdef}), since $N-1$ is the
rank of the ${\mathfrak {su}}(N)$ Lie algebra.
In terms of the $H_{\ell}$ matrices, the electric part of the gauge potential
$A$ (\ref{eq:gaugepot}) takes the form
\begin{equation}
{\mathcal {A}} = -\sum_{\ell=1}^{N-1}h_{\ell} (r) H_{\ell},
\label{eq:Adef}
\end{equation}
in terms of $N-1$ real scalar functions $h_{\ell}(r)$, depending on the radial coordinate $r$ only.

Our ansatz (\ref{eq:Adef}) for the electric part of the gauge potential at first sight looks rather different from that conventionally used in the literature \cite{Winstanley:2008ac,Baxter:2015tda,Kuenzle:1991wa}, although it is equivalent.
Usually the $N$ diagonal entries of the traceless matrix ${\mathcal {A}}$ are denoted by $i\alpha _{k}(r)$, $k=1,\ldots N$ (where the $\alpha _{k}(r)$ are real functions of $r$), and written in terms of $N-1$ real quantities ${\mathcal {E}}_{j}(r)$ as follows \cite{Baxter:2015tda,Kuenzle:1991wa}:
\begin{equation}
\alpha _{k}(r) = - \frac {1}{N} \sum_{\ell=1}^{k-1} \ell {\mathcal {E}}_{\ell }(r) + \sum _{\ell=k}^{N-1} \left( 1 - \frac {\ell }{N} \right) {\mathcal {E}}_{\ell }(r),
\end{equation}
so that
\begin{equation}
{\mathcal {E}}_{k}(r) = \alpha _{k}(r) - \alpha _{k+1}(r).
\end{equation}
From (\ref{eq:Adef}), the $\alpha _{k}(r)$ can be written in terms of the $h_{\ell }(r)$ as:
\begin{equation}
\alpha _{k}(r) = -{\sqrt {\frac {k-1}{2k}}} h_{k-1}(r) + \sum _{\ell = k}^{N-1} \frac {h_{\ell }(r)}{{\sqrt {2\ell \left( \ell +1 \right) }}},
\end{equation}
and consequently
\begin{equation}
{\mathcal {E}}_{k}(r) = {\sqrt {\frac {k+1}{2k}}} h_{k}(r) - {\sqrt {\frac {k-1}{2k}}} h_{k-1}(r).
\label{eq:calEdef}
\end{equation}
We write the relation (\ref{eq:calEdef}) between the $N-1$ functions $h_{\ell }(r)$ and the $N-1$ quantities ${\mathcal {E}}_{k}(r)$ as
\begin{equation}
{\bmath {\mathcal {E}}} = {\mathcal {F}}_{N-1} {\bmath {h}},
\label{eq:calEh}
\end{equation}
where ${\bmath {\mathcal {E}}} = \left( {\mathcal {E}}_{1}, \ldots , {\mathcal {E}}_{N-1} \right)^{T}$, ${\bmath {h}}=\left( h_{1},\ldots ,h_{N-1}\right) ^{T}$ and ${\mathcal {F}}_{N-1}$ is the lower-triangular $\left( N - 1 \right) \times \left( N-1 \right) $ matrix with entries
\begin{widetext}
\begin{equation}
{\mathcal {F}}_{N-1} = \left(
\begin{array}{ccccc}
{\sqrt {2/(2\times 1)}} & 0 & 0 & \cdots & 0 \\
-{\sqrt {1/(2\times 2)}} & {\sqrt {3/(2\times 2)}} & 0 & \cdots & 0 \\
0 & -{\sqrt {2/(2\times 3)}} & {\sqrt {4/(2\times 3)}} & \cdots & 0 \\
\vdots & \vdots & \vdots & \cdots & \vdots \\
0 & 0 & 0 & \cdots & 0\\
0 & 0 & 0 & \cdots & {\sqrt {N/(2\times (N-1))}}
\end{array}
\right) .
\label{eq:calFdef}
\end{equation}
\end{widetext}
By inverting ${\mathcal {F}}_{N-1}$ (\ref{eq:calFdef}), we can write the $h_{\ell }(r)$ in terms of the ${\mathcal {E}}_{k}(r)$:
\begin{equation}
{\bmath {h}} = {\mathcal {F}}_{N-1}^{-1} {\bmath {\mathcal {E}}}.
\end{equation}

For the magnetic part of the gauge field potential (\ref{eq:gaugepot}), we now assume that all the functions $\omega _{j}(r)$ (\ref{eq:Cj}) in the matrix $C$
are nonzero.
In this case one of the Yang-Mills equations reduces to $\gamma _{j}(r)\equiv 0$ for all
$j=1, \ldots , N-1$ and $r$ \cite{Kuenzle:1991wa}.
Our ansatz for the gauge potential (\ref{eq:gaugepot}) then reduces to:
\begin{eqnarray}
A & = & {\mathcal {A}} \, dt
+\frac {1}{2} \left( C-C^{H} \right) d\theta
\nonumber \\ & &
-\frac {i}{2} \left[ \left( C+C^{H} \right) \sin \theta
+D\cos \theta \right] d\phi ,
\label{eq:gaugepotsimp}
\end{eqnarray}
and the nonzero entries in the matrix $C$ are now simply
\begin{equation}
C_{j, j+1} = \omega _{j}(r),
\qquad
j = 1,  \ldots , N-1.
\label{eq:Cjsimp}
\end{equation}
The above ansatze for the matrices ${\mathcal {A}}$ and $C$ appearing in the electric and magnetic parts of the gauge field respectively, are related to the expansion of the ${\mathfrak {su}}(N)$ gauge field in terms of simple roots (such an expansion is well-known in the construction of the ${\mathfrak {su}}(N)$ monopole, see, for example, \cite{Shnir:2005vq} for a review).
The matrix ${\mathcal {A}}$ (\ref{eq:Adef}) is a linear combination of the generators ${\vec {\Upsilon }}_{\ell }.{\vec {H}}$, where the ${\vec {\Upsilon }}_{\ell }$ are the positive roots of ${\mathfrak {su}}(N)$, while the matrix $C$ with entries given by (\ref{eq:Cjsimp}) is a linear combination of the generators $E_{{\vec {\Upsilon }}_{\ell }}$ (the raising operators) corresponding to the simple roots of ${\mathfrak {su}}(N)$.

In summary, the gauge field is described by $2\left( N-1 \right)$ functions of $r$:
the $N-1$ electric gauge functions $h_{\ell}(r)$ and the $N-1$ magnetic
gauge functions $\omega _{j}(r)$.

\subsubsection{Field equations}
\label{sec:eqns}

The components of the Yang-Mills gauge field are given in terms of
the gauge potential components (\ref{eq:gaugepotsimp}) as follows:
\begin{equation}
F_{\mu \nu } = \partial _{\mu } A_{\nu } - \partial _{\nu } A_{\mu }
+ \left[ A_{\mu }, A_{\nu } \right] ,
\label{eq:gaugefield}
\end{equation}
when the gauge coupling constant $g=1$.
The Yang-Mills equations take the form
\begin{equation}
\nabla _{\mu } F^{\mu \nu } + \left[ A_{\mu }, F^{\mu \nu } \right] =0.
\label{eq:YMe}
\end{equation}
The stress-energy tensor of the Yang-Mills field is
\begin{equation}
T_{\mu \nu } = {\mathrm {Tr}} \left[ F_{\mu \alpha } F_{\nu }{}^{\alpha }
- \frac {1}{4} g_{\mu \nu } F_{\alpha \beta } F^{\alpha \beta } \right] ,
\label{eq:stress}
\end{equation}
where ${\mathrm {Tr}}$ denotes a Lie algebra trace.
The stress-energy tensor $T_{\mu \nu }$ acts as the source in the Einstein equations:
\begin{equation}
R_{\mu \nu } - \frac {1}{2} g_{\mu \nu } R + \Lambda g_{\mu \nu } =
2 T_{\mu \nu } ,
\label{eq:Ee}
\end{equation}
since we are using units in which $4\pi G=1$.

\begin{widetext}
The Yang-Mills equations (\ref{eq:YMe}) for the gauge field with potential
(\ref{eq:gaugepotsimp}) take the form (a prime $'$ denotes differentiation with
respect to the radial coordinate $r$):
\begin{subequations}
\label{eq:YMefinal}
\begin{eqnarray}
h_{k} '' & = &
h_{k}' \left( \frac {\sigma'}{\sigma} - \frac{2}{r} \right)
+ \frac {{\sqrt {k+1}}}{k} \frac{\omega _{k}^{2}}{\mu r^{2}}
\left( h_{k} {\sqrt {k+1}}  - h_{k-1} {\sqrt {k-1}} \right)
+\frac {{\sqrt {k}}}{k+1} \frac {\omega_{k+1}^{2}}{\mu r^{2}}
\left( h_{k} {\sqrt {k}}
- h_{k+1} {\sqrt {k+2}} \right) ,
\label{eq:hkeqn}
\\
0 & = &
\omega _{k}''
+ \omega _{k}' \left( \frac {\sigma '}{\sigma } + \frac {\mu '}{\mu } \right)
+ \frac{\omega _{k}}{\sigma ^{2} \mu ^{2}}
\left( h_{k} {\sqrt {\frac {k+1}{2k}}}
- h_{k-1} {\sqrt {\frac {k-1}{2k}}} \right)^2
+ \frac {\omega _{k}}{2\mu r^{2}}
\left( 2 - 2\omega _{k}^{2}
+ \omega _{k-1}^{2} + \omega _{k+1}^{2} \right) ,
\label{eq:omegakeqn}
\end{eqnarray}
\end{subequations}
for $k=1,\ldots , N-1$.
The Einstein equations (\ref{eq:Ee}) give two equations for the metric functions
$m(r)$ and $\sigma (r)$:
\begin{subequations}
\label{eq:Eefinal}
\begin{eqnarray}
m ' & = &
\sum_{k=1}^{N-1} \left[
 \frac {r^{2} h_{k}'^{2}}{2\sigma ^{2}}
 +
\frac {\omega _{k}^{2}}{\sigma ^{2} \mu }
\left( h_{k} {\sqrt{ \frac {k+1}{2k}}} - h_{k-1} {\sqrt {\frac {k-1}{2k}}}
\right) ^{2}
\right]
+ \sum_{k=1}^{N-1} \left[
\mu \omega _{k}'^{2}
+ \frac {k(k+1)}{4r^{2}} \left(
1 -  \frac {\omega_{k}^{2}}{k}
+ \frac {\omega _{k+1}^{2}}{k+1} \right) ^{2}
\right] ,
\label{eq:mprime}
\\
\sigma ' & = &
\sum_{k=1}^{N-1} \left[
\frac {2\sigma \omega _{k}'^{2}}{r}
+
\frac {2\omega _{k}^{2}}{\sigma \mu ^{2} r}
\left( h_{k} {\sqrt {\frac {k+1}{2k}}} - h_{k-1} {\sqrt {\frac {k-1}{2k}}}
\right) ^{2}
\right] .
\label{eq:sigmaprime}
\end{eqnarray}
\end{subequations}
Setting $h_{k}(r) \equiv 0$ for all $k$, the field equations
(\ref{eq:YMefinal}--\ref{eq:Eefinal}) reduce to the purely magnetic
field equations studied in \cite{Baxter:2007at,Baxter:2008pi,Baxter:2015gfa}.
\end{widetext}

\subsection{Boundary conditions}
\label{sec:boundary}

The field equations (\ref{eq:YMefinal}--\ref{eq:Eefinal}) are singular at the origin
$r=0$ (relevant for soliton solutions), the black hole event horizon $r=r_{h}$ (if
there is one) and  infinity $r\rightarrow \infty $.
We therefore need to specify appropriate boundary conditions at each of these points.
As with the purely magnetic solutions \cite{Baxter:2008pi} the
boundary conditions at the origin are the most complex.
Local existence of solutions of the field equations near the singular points $r=0$, $r=r_{h}$ and $r\rightarrow \infty $ is proven in \cite{Baxter:2015tda},
using a different representation of the electric part of the gauge field potential.
In this section we cast the results of \cite{Baxter:2015tda} into our notation for completeness.

\subsubsection{Infinity}
\label{sec:infinity}

We assume that the space-time is asymptotically adS and that
 the field variables have regular Taylor series expansions about
$r\rightarrow \infty $:
\begin{eqnarray}
m(r) & = & M + \frac {m_{1}}{r}
+ {\mathcal {O}}\left(\frac {1}{r^{2}}\right),
\nonumber \\
\sigma (r) & = & 1 + \frac {\sigma _{1}}{r}
+ \frac {\sigma _{2}}{r^{2}}
+ \frac {\sigma _{3}}{r^{3}}
+ \frac {\sigma _{4}}{r^{4}}
+ {\mathcal{O}} \left( \frac {1}{r^{5}} \right),
\nonumber \\
\omega _{k} (r) & = &
\omega _{k,\infty } + \frac {c_{k,1}}{r}
+ {\mathcal {O}} \left( \frac {1}{r^{2}} \right) ,
\nonumber \\
h_{k} (r) & = & h_{k,\infty } + \frac {d_{k,1}}{r}
+ {\mathcal{O}} \left( \frac{1}{r^{2}} \right) .
\label{eq:infinity}
\end{eqnarray}
Substituting the expressions (\ref{eq:infinity}) into the field equations
(\ref{eq:YMefinal}--\ref{eq:Eefinal}), we find that
$h_{k,\infty }$, $d_{k,1}$, $\omega _{k,\infty }$ and $c _{k,1}$ are free
parameters,
\begin{equation}
\sigma _{1} = \sigma _{2} = \sigma _{3} = 0,
\label{eq:sigmainfinity}
\end{equation}
and that $m_{1}$ and $\sigma _{4}$ are given by (cf.~\cite{Baxter:2015tda})
\begin{widetext}
\begin{eqnarray}
m_{1} & = &
-\sum _{k=1}^{N-1}
\left[ \frac {k(k+1)}{4}
\left( 1 - \frac {\omega _{k, \infty}^{2}}{k}
+ \frac {\omega _{k+1, \infty }^{2}}{k+1} \right) ^{2}
+ \frac {d_{k,1}^{2}}{2} + \frac {c_{k,1}^{2}}{L^{2}}
-  L^{2} \omega _{k,\infty}^{2}
\left( h_{k,\infty } {\sqrt {\frac {k+1}{2k}}}
- h_{k-1,\infty } {\sqrt {\frac {k-1}{2k}}}
\right) ^{2} \right] ,
\nonumber
\\
\sigma _{4} & = &
- \frac {1}{2} \sum_{k=1}^{N-1}
\left[ L^{4} \omega _{k,\infty }^{2}
\left( h_{k,\infty } {\sqrt {\frac {k+1}{2k}}}
 - h_{k-1,\infty } {\sqrt {\frac {k-1}{2k}}} \right) ^{2} + c_{k,1}^{2} \right],
\label{eq:m1sigma4infinity}
\end{eqnarray}
where $L={\sqrt {-3/\Lambda }}$ is the adS radius of curvature.
\end{widetext}

\subsubsection{Event horizon}
\label{sec:eventhorizon}

We assume that there is a regular, nonextremal black hole event horizon at $r=r_{h}$.
At $r=r_{h}$, the metric function $\mu (r)$ (\ref{eq:mu}) has a single zero, so
that
\begin{equation}
m(r_{h}) = \frac {r_{h}}{2} - \frac {\Lambda r_{h}^{3}}{6} .
\label{eq:mrh}
\end{equation}
To avoid a singularity at the event horizon, it must be the case that the
electric gauge functions $h_{k}(r)$ vanish at $r=r_{h}$.
We therefore assume the following Taylor series expansions near the event horizon:
\begin{eqnarray}
m(r) & = & m(r_{h}) + m'(r_{h})\left( r-r_{h}\right)
+ {\mathcal {O}}\left( r-r_{h}\right) ^{2},
\nonumber \\
\sigma (r) & = & \sigma (r_{h}) + \sigma '(r_{h}) \left( r-r_{h} \right)
+ {\mathcal {O}} \left( r-r_{h} \right) ^{2},
\nonumber\\
\omega _{k}(r) & = & \omega _k (r_{h}) + \omega _{k}'(r_{h})\left( r-r_{h}\right)
+ {\mathcal {O}} \left( r-r_{h}\right) ^{2},
\nonumber \\
h_{k}(r) & = & h_{k}'(r_{h})\left( r-r_{h}\right)
+ {\mathcal {O}} \left( r-r_{h} \right) ^{2}.
\label{eq:horizon}
\end{eqnarray}

From the field equations, we find that $\sigma (r_{h})$,
$h_{k}'(r_{h})$ and $\omega _{k}(r_{h})$
are free parameters and that
$m'(r_{h})$, $\sigma '(r_{h})$ and $\omega _{k}'(r_{h})$ are given in terms of them as
follows (cf.~\cite{Baxter:2015tda}):
\begin{widetext}
\begin{eqnarray}
m'(r_{h}) &  =  &
\sum _{k=1}^{N-1} \left[
\frac {r_{h}^{2} h_{k}'(r_{h})^{2}}{2\sigma (r_{h})^{2}}
+ \frac {k(k+1)}{4r_{h}^{2}}\left(
1 - \frac {\omega _{k}(r_{h})^{2}}{k}
+ \frac {\omega _{k+1}(r_{h})^{2}}{k+1} \right) ^{2}
\right] ,
\nonumber \\
\sigma '(r_{h}) &  =  &
2\sum _{k=1}^{N-1} \left[
\frac {\omega _{k}(r_{h})^{2}}{\sigma (r_{h}) \mu '(r_{h})^{2} r_{h}}
\left( h_{k}'(r_{h}) {\sqrt {\frac {k+1}{2k}}}
- h_{k-1}'(r_{h}) {\sqrt {\frac {k-1}{2k}}} \right) ^{2}
\right]
+ 2\sum_{k=1}^{N-1} \frac {\sigma (r_{h}) \omega _{k}'(r_{h})^{2}}{r_{h}} ,
\nonumber \\
\omega _{k}'(r_{h}) &  =  &
\frac{ \omega _{k}(r_{h})}{\mu '(r_{h}) r_{h}^{2}}
\left( \omega _{k}(r_{h})^{2} - 1 - \frac{1}{2}
\left[ \omega _{k-1}(r_{h})^{2} + \omega _{k+1}(r_{h})^{2} \right] \right) .
\label{eq:horexp}
\end{eqnarray}
\end{widetext}

The value of $\sigma (r_{h})$ is fixed in practice by the requirement that the metric
function $\sigma (r)$ approaches unity as $r\rightarrow \infty $.
This leaves the $2\left( N - 1\right) $ free parameters $h_{k}'(r_{h})$ and
$\omega _{k}(r_{h})$ for $k=1,\ldots , N-1$, whose values are restricted by the
requirement
\begin{equation}
\mu '(r_{h}) = \frac{1}{r_{h}} - \frac{2m'(r_{h})}{r_{h}} - \Lambda r_{h} > 0 ,
\label{eq:murh}
\end{equation}
which is needed for a regular nonextremal horizon at $r=r_{h}$.

\subsubsection{Origin}
\label{sec:origin}

In the purely magnetic case \cite{Baxter:2007at,Baxter:2008pi}, the boundary conditions
for the ${\mathfrak {su}}(N)$ gauge potential near the origin are rather complicated,
with a power series expansion up to ${\mathcal {O}}(r^{N+1})$ necessary in order to
completely specify the gauge field functions.
It is no surprise that the addition of a nontrivial electric part to the gauge field potential only complicates matters further \cite{Baxter:2015tda}.

We begin by assuming regular Taylor series expansions for all field variables in
a neighbourhood of the origin $r=0$:
\begin{eqnarray}
m(r) & = & m_{0} + m_{1} r + m_{2} r^{2} + m_{3}r^{3} + {\mathcal {O}}(r^{4}),
\nonumber \\
\sigma (r) & = & \sigma _{0} + \sigma _{1}r + \sigma _{2}r^{2} +
{\mathcal {O}}(r^{3}) ,
\nonumber \\
\omega _{k} (r) & = & \omega _{k,0} + \omega _{k,1} r + \omega _{k,2}r^{2}
+ \omega _{k,3} r^{3} + {\mathcal {O}}(r^{4}),
\nonumber \\
h_{k}(r) & =& h_{k,0} + h_{k,1}r + h_{k,2}r^{2} + h_{k,3}r^{3}
+ {\mathcal {O}} (r^{4}).
\label{eq:origin}
\end{eqnarray}
The constant $\sigma _{0}$ must be nonzero in order for the metric (\ref{eq:metric})
to be regular at the origin.
It is otherwise arbitrary as far as the expansions near the origin are concerned, and
will be fixed in practice by the requirement that $\sigma \rightarrow 1$ as
$r\rightarrow \infty $.
Regularity of the field equations (\ref{eq:YMefinal}--\ref{eq:Eefinal}), metric
(\ref{eq:metric}) and curvature at $r=0$ gives \cite{Baxter:2015tda}
\begin{equation}
m_{0}=m_{1}=m_{2}=0, \quad
\sigma _{1}=0, \quad
\omega _{k,1} = 0, \quad
h_{k,0} =0.
\label{eq:originconds1}
\end{equation}
From the equation for $m'(r)$ (\ref{eq:mprime}), we also have
\begin{equation}
\sum _{k=1}^{N-1}
k\left( k +1 \right) \left[ 1 - \frac {\omega _{k,0}^{2}}{k}
+ \frac {\omega _{k+1,0}^{2}}{k+1} \right] ^{2} =0,
\label{eq:originconds2}
\end{equation}
which is solved by
\begin{equation}
\omega _{k,0} = \pm {\sqrt {k\left( N - k\right) }}.
\label{eq:omegaorigin}
\end{equation}
The field equations (\ref{eq:YMefinal}--\ref{eq:Eefinal}) are unchanged by
the transformation $\omega _{k}(r) \rightarrow - \omega _{k}(r)$, for each
$k$ separately.
Therefore we take the positive sign in (\ref{eq:omegaorigin}) without
loss of generality.

We consider next the magnetic gauge field functions $\omega _{k}(r)$.
The coupling between these and the electric gauge field functions $h_{k}(r)$
in the Yang-Mills equation for $\omega _{k}(r)$ (\ref{eq:omegakeqn}) does not affect
the first two terms in the expansion of this equation near $r=0$, because
$h_{k}(r)={\mathcal {O}}(r)$ as $r\rightarrow 0$.
Following \cite{Baxter:2008pi,Baxter:2015tda}, we define two vectors
${\bmath {\omega }}_{2} = \left( \omega _{1,2} , \omega _{2,2}, \ldots
, \omega _{N-1,2} \right) ^{T}$ and
${\bmath {\omega }}_{3} = \left( \omega _{1,3} , \omega _{2,3}, \ldots
, \omega _{N-1,3} \right) ^{T}$.
The first two terms in the expansion about $r=0$ of the
Yang-Mills equation (\ref{eq:omegakeqn}) then give the following equations for
${\bmath {\omega }}_{2}$ and ${\bmath {\omega }}_{3}$:
\begin{equation}
{\mathcal {M}}_{N-1} {\bmath {\omega }}_{2} = 2{\bmath {\omega }}_{2} ,
\qquad
{\mathcal {M}}_{N-1} {\bmath {\omega }}_{3} = 6 {\bmath {\omega }}_{3},
\label{eq:omega23}
\end{equation}
where ${\mathcal {M}}_{N-1}$ is the $\left( N-1 \right) \times \left( N-1 \right) $
matrix \cite{Baxter:2008pi,Baxter:2015tda,Kuenzle:1994ru}
\begin{widetext}
\begin{equation}
{\mathcal {M}}_{N-1} = \left(
\begin{array}{ccccc}
2(N-1)&-\sqrt{(N-1)2(N-2)}&0&\cdots& 0 \\
-\sqrt{(N-1)2(N-2)}&2\times 2(N-2)&-\sqrt{2(N-2)3(N-3)}&\cdots&0 \\
0&-\sqrt{2(N-2)3(N-3)}&2\times 3(N-3)&\cdots & 0\\
\vdots&\vdots&\vdots&\ddots & \vdots\\
0 & 0 & 0 & \cdots  & -\sqrt{(N-1)2(N-2)}\\
0 & 0 & 0 & \cdots  & 2(N-1)
\end{array}
\right) .
\label{eq:calM}
\end{equation}
\end{widetext}
As discussed in \cite{Kuenzle:1994ru}, the eigenvalues of the matrix
${\mathcal {M}}_{N-1}$ are
\begin{equation}
{\sf {E}}_{j} = j \left( j + 1 \right) ,
\qquad j = 1, \ldots , N-1 ,
\label{eq:eigenvalues}
\end{equation}
and the eigenvectors have a complicated form involving Hahn polynomials \cite{Kuenzle:1994ru, Hahn}.
Writing a basis of normalized eigenvectors of ${\mathcal {M}}_{N-1}$ as
${\bmath {u}}_{1}, {\bmath {u}}_{2}, \ldots , {\bmath {u}}_{N-1}$
(where ${\bmath {u}}_{j}$ is the eigenvector having eigenvalue ${\sf {E}}_{j}$
(\ref{eq:eigenvalues})), we have, from (\ref{eq:omega23}),
\begin{equation}
{\bmath {\omega }}_{2} = b_{1} {\bmath {u}}_{1},
\qquad
{\bmath {\omega }}_{3} = b_{2} {\bmath {u}}_{2},
\label{eq:omega23sol}
\end{equation}
where $b_{1}$ and $b_{2}$ are arbitrary constants.

We proceed in a similar way for the electric gauge field functions $h_{k}(r)$,
defining two vectors
${\bmath {h}}_{1} = \left( h_{1,1} , h_{2,1}, \ldots
, h_{N-1,1} \right) ^{T}$ and
${\bmath {h}}_{2} = \left( h_{1,2} , h_{2,2}, \ldots
, h_{N-1,2} \right) ^{T}$.
As with the magnetic gauge field functions, the coupling between the
$h_{k}(r)$ and  $\omega _{k}(r)$ in the Yang-Mills equation (\ref{eq:hkeqn})
does not affect the first two terms in the expansion of this equation about $r=0$,
which then give the following two equations:
\begin{equation}
{\mathcal {N}}_{N-1} {\bmath {h}}_{1} = 2{\bmath {h}}_{1}, \qquad
{\mathcal {N}}_{N-1} {\bmath {h}}_{2} = 6{\bmath {h}}_{2},
\label{eq:h23}
\end{equation}
where ${\mathcal {N}}_{N-1}$ is the $\left( N-1 \right) \times \left( N-1 \right) $
matrix
\begin{widetext}
\begin{equation}
{\mathcal {N}}_{N-1} = \left(
\begin{array}{ccccc}
2(N-1)+(N-2) & -(N-2){\sqrt {3}} & 0 &\cdots & 0 \\
-(N-2){\sqrt {3}} & 2\times 2(N-2)+(N-4) & -(N-3){\sqrt {2\times 4}} &\cdots&0 \\
0& -(N-3){\sqrt {2\times 4}} & 2\times 3(N-3)+(N-6) &\cdots & 0\\
\vdots&\vdots&\vdots&\ddots & \vdots\\
0 & 0 & 0 & \cdots  & - {\sqrt {(N-2) N}} \\
0 & 0 & 0 & \cdots  & 2(N-1)-(N-2)
\end{array}
\right) .
\label{eq:calN}
\end{equation}
\end{widetext}

In \cite{Baxter:2015tda}, an expansion similar to (\ref{eq:origin}) is performed for the alternative electric gauge field functions ${\mathcal {E}}_{k}$, defined in terms of the $h_{\ell }$ functions that we use here by (\ref{eq:calEdef}, \ref{eq:calEh}):
\begin{equation}
{\mathcal {E}}_{k}(r)  = {\mathcal {E}}_{k,0} + {\mathcal {E}}_{k,1}r + {\mathcal {E}}_{k,2}r^{2} + {\mathcal {E}}_{k,3}r^{3}
+ {\mathcal {O}} (r^{4}).
\end{equation}
Again we have ${\mathcal {E}}_{k,0}=0$ for all $k$, and the vectors ${\bmath {\mathcal {E}}}_{1}=\left( {\mathcal {E}}_{1,1},  {\mathcal {E}}_{2,1} \ldots ,
{\mathcal {E}}_{N-1,1}\right) ^{T}$ and ${\bmath {\mathcal {E}}}_{2}=\left( {\mathcal {E}}_{1,2}, {\mathcal {E}}_{2,2} \ldots ,
{\mathcal {E}}_{N-1,2}\right) ^{T}$ satisfy the equations \cite{Baxter:2015tda}
\begin{equation}
{\mathcal {M}}_{N-1} {\bmath {\mathcal {E}}}_{1} = 2 {\bmath {\mathcal {E}}}_{1}, \qquad
{\mathcal {M}}_{N-1} {\bmath {\mathcal {E}}}_{2} = 6 {\bmath {\mathcal {E}}}_{2} .
\end{equation}
Using the relationship between ${\bmath {h}}$ and ${\bmath {\mathcal {E}}}
$ in (\ref{eq:calEh}), it is clear that the matrices ${\mathcal {M}}_{N-1}$ (\ref{eq:calM}) and ${\mathcal {N}}_{N-1}$  (\ref{eq:calN}) are related by
\begin{equation}
{\mathcal {N}}_{N-1} = {\mathcal {F}}_{N-1}^{-1} {\mathcal {M}}_{N-1} {\mathcal {F}}_{N-1},
\end{equation}
and therefore the matrices ${\mathcal {M}}_{N-1}$ and ${\mathcal {N}}_{N-1}$ have the same eigenvalues ${\sf {E}}_{j}$ (\ref{eq:eigenvalues}).
In particular, we have, from
(\ref{eq:h23}),
\begin{equation}
{\bmath {h}}_{1} = g_{1} {\bmath {v}}_{1},
\qquad
{\bmath {h}}_{2} = g_{2} {\bmath {v}}_{2},
\label{eq:h23sol}
\end{equation}
where $g_{1}$ and $g_{2}$ are arbitrary constants and the vectors ${\bmath {v}}_{i}$, $i=1, 2$, are eigenvectors of ${\mathcal {N}}_{N-1}$, given by
\begin{equation}
{\bmath {v}}_{i} \propto {\mathcal {F}}_{N-1}^{-1} {\bmath {u}}_{i}, \qquad i=1,2 ,
\end{equation}
where an overall multiplicative constant is required so that the ${\bmath {v}}_{i}$ are normalized.

Once we have found the eigenvectors ${\bmath {\omega }}_{2}$, ${\bmath {\omega }}_{3}$,
${\bmath {h}}_{1}$ and ${\bmath {h}}_{2}$, the first two terms in the Einstein equations
(\ref{eq:mprime}, \ref{eq:sigmaprime}) give the values of
$m_{3}, m_{4}, \sigma _{2}$ and $\sigma _{3}$ (which also depend on the
cosmological constant $\Lambda $) \cite{Baxter:2015tda}.

From the above analysis, the expansion of the $h_{k}(r)$ to order $r$ and $\omega _{k}(r)$ functions
to order $r^{2}$ depends on just two arbitrary constants, $b_{1}$ (\ref{eq:omega23sol})
and $g_{1}$ (\ref{eq:h23sol}), while the expansion to next order in $r$ (that is, to order $r^{2}$ for the electric gauge field functions $h_{k}(r)$ and order $r^{3}$ for the magnetic gauge field functions $\omega _{k}(r)$) adds
a further two arbitrary constants, $b_{2}$ and $g_{2}$.
It is shown in Proposition 8 in \cite{Baxter:2015tda}, in analogy with the purely magnetic case \cite{Baxter:2008pi},
that a total of $2\left( N - 1 \right)$ arbitrary constants are required to completely specify the gauge field functions
in a neighbourhood of the origin.
Each additional power of $r$ in the expansion of the $h_{k}(r)$
and $\omega _{k}(r)$  depends on just two further arbitrary constants, one for the
$h_{k}(r)$ and one for the $\omega _{k}(r)$.

To see this, define the vectors
\begin{eqnarray}
{\bmath {h}}_{j} & = & \left( h_{1,j}, h_{2,j},\ldots , h_{N-1,j} \right) ^{T},
\nonumber
\\
{\bmath {\omega }}_{j} & = &
\left( \omega _{1,j}, \omega _{2,j} , \ldots , \omega _{N-1,j} \right) ^{T}.
\label{eq:hjomegaj}
\end{eqnarray}
Examination of the appropriate term in the expansion of the relevant Yang-Mills
equation (\ref{eq:YMefinal}) shows that the vectors
${\bmath {h}}_{j}$, ${\bmath {\omega }}_{j}$ satisfy equations of the form
\begin{eqnarray}
\left[ {\mathcal {M}}_{N-1} - j \left( j + 1 \right) \right]
{\bmath {\omega }}_{j+1} & = & {\bmath {p}}_{j+1} ,
\nonumber \\
\left[ {\mathcal {N}}_{N-1} - j \left( j + 1 \right) \right]
{\bmath {h}}_{j} & = & {\bmath {q}}_{j+1},
\label{eq:higherorigin}
\end{eqnarray}
where ${\bmath {p}}_{j+1}$ and ${\bmath {q}}_{j+1}$ are complicated
vectors depending on ${\bmath {h}}_{2}, {\bmath {h}}_{3}, \ldots , {\bmath {h}}_{j-1}$,
${\bmath {\omega }}_{2}, {\bmath {\omega }}_{3}, \ldots
{\bmath {\omega }}_{j}$, $m_{3}, m_{4}, \ldots m_{j}$, $\sigma _{2}, \sigma _{3}, \ldots , \sigma _{j}$, whose detailed form can be found in \cite{Baxter:2015tda}.
The solutions of (\ref{eq:higherorigin}) are \cite{Baxter:2015tda}:
\begin{eqnarray}
{\bmath {h}}_{j} & = & g_{j}{\bmath {v}}_{j}
+ {\tilde {\bmath {v}}}_{j+1},
\nonumber \\
{\bmath {\omega }}_{j+1} & = & b_{j} {\bmath {u}}_{j} + {\tilde {\bmath {u}}}_{j+1},
\label{eq:highersol}
\end{eqnarray}
where $b_{j}$ and $g_{j}$ are arbitrary constants.
In (\ref{eq:highersol}), the vectors ${\tilde {\bmath {u}}}_{j+1}$ and
${\tilde {\bmath {v}}}_{j+1}$ are particular solutions of (\ref{eq:higherorigin}).
It is shown in \cite{Baxter:2015tda} that these particular solutions can be chosen such that ${\tilde {\bmath {u}}}_{j+1}$
and ${\tilde {\bmath {v}}}_{j+1}$
are linear combinations of ${\bmath {u}}_{1}, {\bmath {u}}_{2}, \ldots ,
{\bmath {u}}_{j-1}, {\bmath {v}}_{1}, {\bmath {v}}_{2}, \ldots , {\bmath {v}}_{j-1}$.

We can therefore write the electric and magnetic gauge field functions in the following vectorial form, where ${\bmath {\omega }}(r) = \left( \omega _{1}(r), \omega _{2}(r), \ldots , \omega _{N-1}(r) \right) ^{T}$ and
${\bmath {h}}(r) = \left( h_{1}(r), h_{2}(r), \ldots , h_{N-1}(r) \right) ^{T}$:
\begin{eqnarray}
{\bmath {\omega }}(r) & = &
{\bmath {\omega }}_{0} + \sum _{k=1}^{N-1} \beta _{k}(r) {\bmath {u}}_{k},
\nonumber \\
{\bmath {h}}(r) & = & \sum _{k=1}^{N-1} \gamma _{k}(r) {\bmath {v}}_{k},
\label{eq:betagamma}
\end{eqnarray}
where
\begin{equation}
{\bmath {\omega }}_{0} = \left( {\sqrt {N-1}}, {\sqrt {2\left( N - 2 \right) }}, \ldots, {\sqrt {N-1}} \right) ^{T},
\end{equation}
and the $\beta _{k}(r)$ and $\gamma _{k}(r)$ functions have the following behaviour near the origin:
\begin{eqnarray}
\beta _{k} (r) & = &
b_{k}r^{k+1} + {\mathcal {O}}(r^{k+2}),
\nonumber \\
\gamma _{k}(r) & = &
g_{k}r^{k} + {\mathcal {O}}(r^{k+1}) .
\label{eq:betagammaorigin}
\end{eqnarray}
As is explained in more detail in the next section, in our numerical procedure we integrate the field equations for the $\beta _{k}(r)$ and $\gamma _{k}(r)$ functions rather than $h_{k}(r)$ and $\omega _{k}(r)$, to improve numerical accuracy.

In this paper we present dyonic solutions for the $N=2$ and $N=3$ cases
only, so it is sufficient for our purposes to find ${\bmath {h}}_{2}$,
${\bmath {h}}_{3}$, ${\bmath {\omega }}_{2}$ and ${\bmath {\omega }}_{3}$.
We do not need to consider the complicated vectors ${\bmath {p}}_{j+1}$, ${\bmath {q}}_{j+1}$ (\ref{eq:higherorigin}) or ${\tilde {{\bmath {u}}}}_{j+1}$,
${\tilde {{\bmath {v}}}}_{j+1}$ (\ref{eq:highersol}).

\section{Solutions}
\label{sec:solutions}

We now present our new soliton and black hole solutions of the field equations (\ref{eq:YMefinal}--\ref{eq:Eefinal}).
These field equations have a number of trivial solutions, which we describe first in Sec.~\ref{sec:trivial}, before discussing our numerical method in Sec.~\ref{sec:numerics} and the nontrivial solutions for ${\mathfrak {su}}(2)$ and ${\mathfrak {su}}(3)$ gauge groups in Secs.~\ref{sec:su2} and \ref{sec:su3} respectively.

\subsection{Trivial solutions}
\label{sec:trivial}

The first trivial solution arises on setting the electric gauge functions $h_{k}(r)\equiv 0$ for $k=1,\ldots , N-1$,
and the magnetic gauge functions $\omega _{k}(r)$ to be the following constants:
\begin{equation}
\omega _{k}(r) \equiv \pm {\sqrt {k\left( N - k \right) }},
\qquad
k=1, \ldots , N-1 .
\end{equation}
The metric functions $m(r)$ and $\sigma (r)$ are then constants.  Setting
$\sigma (r) \equiv 1$ without loss of generality gives the Schwarzschild-adS metric
with $m(r)\equiv M$.

The second trivial solution is Reissner-Nordstr\"om-adS.
The metric function $\sigma (r) \equiv 1$ and $\mu (r)$ takes the form
\begin{equation}
\mu _{RN} (r) = 1 - \frac {2M}{r} + \frac {Q^{2}}{r^{2}} - \frac {\Lambda r^{2}}{3} ,
\label{eq:muRN}
\end{equation}
where the mass $M$ and charge $Q$ are constants.
This solution of the field equations arises on setting the magnetic gauge functions
$\omega _{k}(r) \equiv 0$
for all $k=1,\ldots , N-1$, in which case the electric gauge functions are exactly
\begin{equation}
h_{k}(r) = h_{k,\infty } - \frac {d_{k,1}}{r},
\qquad k = 1, \ldots , N-1,
\label{eq:hRN}
\end{equation}
where the $h_{k,\infty }$ and $d_{k,1}$ are constants.
From the Einstein equation (\ref{eq:mprime}), the charge $Q$ is given by
\begin{equation}
Q^{2} =
\frac {1}{6}N \left( N -1 \right) \left( N + 1 \right)
+ \sum _{k=1}^{N-1} d_{k,1}^{2}.
\label{eq:RNcharge}
\end{equation}
The charge $Q$ (\ref{eq:RNcharge}) is an effective charge, with $Q^{2}$ having
two components.
The first, $N \left( N -1 \right) \left( N + 1 \right) /6$, is a magnetic charge,
and the second, $\sum _{k=1}^{N-1} d_{k,1}^{2}$, is an electric charge.
The Reissner-Nordstr\"om-adS solution is therefore dyonic in this case.
Setting all the $d_{k,1}=0$ yields the purely magnetically charged
Reissner-Nordstr\"om-adS solution.
Note that purely electrically charged Reissner-Nordstr\"om-adS is {\em {not}}
a solution of the field equations (\ref{eq:YMefinal}--\ref{eq:Eefinal}) due
to the coupling between the electric gauge field functions $h_{k}(r)$ and
the magnetic gauge field functions $\omega _{k}(r)$.

The third class of trivial solutions is ${\mathfrak {su}}(2)$ embedded solutions, obtained in \cite{Baxter:2015tda} with an alternative parametrization
of the electric part of the gauge field potential.
We start by writing the $N-1$ magnetic gauge functions $\omega _{k}(r)$ in terms of
a single function $\omega (r)$, and the $N-1$ electric gauge functions $h_{k}(r)$
in terms of a single function $h(r)$, as follows:
\begin{equation}
\omega _{k}(r) = A_{k} \omega (r) ,
\qquad
h_{k}(r) = B_{k} h (r),
\label{eq:su2gauge}
\end{equation}
where $A_{k}$ and $B_{k}$ are constants.
The Yang-Mills equations (\ref{eq:YMefinal}) reduce to those
for the ${\mathfrak {su}}(2)$ case with gauge functions $h(r)$ and $\omega (r)$ if the
following conditions hold:
\begin{eqnarray}
1 & = &
\left(
B_{k} {\sqrt {\frac {k+1}{2k}}} - B_{k-1} {\sqrt {\frac {k-1}{2k}}} \right) ^{2}
\nonumber \\ & = &
\frac {1}{2} \left( 2 A_{k}^{2} - A_{k+1}^{2} - A_{k-1}^{2} \right)
\nonumber \\ & = &
\frac {A_{k+1}^{2} {\sqrt {k}}}{2\left( k+1 \right)}
\left( {\sqrt{k}} - \frac {B_{k+1}}{B_{k}} {\sqrt {k+2}} \right)
\nonumber \\ & &
+\frac {A_{k}^{2}{\sqrt{k+1}}}{2k}
\left( {\sqrt {k+1}} - \frac {B_{k-1}}{B_{k}} {\sqrt {k-1}} \right) .
\label{eq:ymsu2conds}
\end{eqnarray}
Substituting (\ref{eq:su2gauge}) into the Einstein equations
(\ref{eq:Eefinal}), we obtain the ${\mathfrak {su}}(2)$
equations if
\begin{eqnarray}
& & \frac {1}{6}N \left( N -1 \right) \left( N +1 \right)
\nonumber \\
& = &
\sum _{k=1}^{N-1} \frac {A_{k}^{2}}{{\sqrt {2k}}} \left(
B_{k} {\sqrt {k+1}} - B_{k-1} {\sqrt {k-1}} \right) ^{2}
\nonumber \\ & = &
\sum _{k=1}^{N-1} A_{k}^{2}
\nonumber \\
& = &
\sum _{k=1}^{N-1} B_{k}^{2},
\label{eq:esu2conds1}
\end{eqnarray}
and
\begin{equation}
1 = \left(
\frac {A_{k}^{2}}{k} - \frac {A_{k+1}^{2}}{k+1}
\right) ^{2} .
\label{eq:esu2conds2}
\end{equation}
The conditions (\ref{eq:ymsu2conds}--\ref{eq:esu2conds2}) are solved by taking
\begin{equation}
A_{k} = {\sqrt {k\left( N - k \right) }},
\qquad
B_{k} = {\sqrt {\frac {1}{2} k \left( k + 1 \right) }}.
\label{eq:ABsu2}
\end{equation}
The values of $A_{k}$ are the same as those used to embed purely magnetic
${\mathfrak {su}}(2)$ solutions into ${\mathfrak {su}}(N)$ EYM \cite{Kuenzle:1994ru}.
Substituting (\ref{eq:su2gauge}, \ref{eq:ABsu2}) into the field equations
(\ref{eq:YMefinal}--\ref{eq:Eefinal}) and defining new rescaled variables as follows \cite{Baxter:2015tda}:
\begin{equation}
R = \lambda _{N}^{-1}r,
\quad
{\tilde{m}} = \lambda _{N}^{-1}m,
\quad
{\tilde{h}} = \lambda _{N} h,
\quad
{\tilde{\Lambda}} = \lambda _{N}^{2} \Lambda,
\label{eq:su2scaling}
\end{equation}
where
\begin{equation}
\lambda _{N} = {\sqrt {\frac {1}{6} N \left( N - 1 \right) \left( N + 1 \right) }} ,
\label{eq:lambdaN}
\end{equation}
gives the ${\mathfrak {su}}(2)$ field equations
\begin{eqnarray}
\frac {d{\tilde{m}}}{dR}
& = &
\frac {R^{2}}{2\sigma ^{2}} \left( \frac {d{\tilde {h}}}{dR} \right) ^{2}
+
\frac {\omega ^{2} {\tilde {h}}^{2}}{\sigma ^{2} \mu}
+ \mu \left( \frac{d\omega }{dR} \right) ^{2}
\nonumber \\ & &
+ \frac {1}{2R^{2}} \left( 1 - \omega ^{2} \right) ^{2},
\nonumber \\
\frac {d\sigma }{dR}
& = &
\frac {2\sigma }{R} \left( \frac {d\omega }{dR}\right) ^{2}
+\frac {2\omega ^{2} {\tilde {h}}^{2}}{R\sigma \mu ^{2}} ,
\nonumber \\
\frac {d^{2}{\tilde {h}}}{dR^{2}}
& = &
\frac {d{\tilde {h}}}{dR}
\left( \frac {1}{\sigma } \frac {d\sigma }{dR} - \frac {2}{R} \right)
+ \frac {2{\tilde {h}} \omega ^{2}}{\mu R^{2}} ,
\nonumber \\
0 			
& = &
\frac {d^{2}\omega }{dR^{2}}
+ \frac {d\omega }{dR}
\left( \frac {1}{\sigma } \frac {d\sigma }{dR} + \frac{1}{\mu }\frac {d\mu }{dR} \right)
\nonumber \\ & &
+ \frac {\omega }{\mu }\left( \frac {{\tilde{h}}^{2}}{\sigma ^{2}\mu }
+ \frac {1}{R^{2}} \left[ 1 - \omega ^{2} \right] \right) ,
\end{eqnarray}
where the magnetic gauge field function $\omega (r)$ and the metric functions $\mu (r)$
and $\sigma (r)$ are not scaled.
Setting ${\tilde {h}}(R)\equiv 0$ gives, as expected, the purely magnetic embedded
${\mathfrak {su}}(2)$ equations.

\subsection{Numerical method}
\label{sec:numerics}

The field equations (\ref{eq:YMefinal}--\ref{eq:Eefinal}) form a set of $2N$ ordinary
differential equations.
Note that, unlike the purely magnetic case \cite{Baxter:2007at}, here the equation for
the metric function $\sigma $ (\ref{eq:sigmaprime}) does not decouple from the other
equations.
This does not complicate the numerical method significantly.
We employ standard ``shooting'' techniques \cite{NR}, using a Bulirsch-Stoer algorithm
in C++ to integrate the ordinary differential equations.
The field equations are singular at the origin or a black hole event horizon.

For black hole solutions, we start our integration just outside the event horizon,
at $r=r_{h}+10^{-7}$, using the expansions (\ref{eq:horizon}) as initial conditions.
We then integrate outwards with $r$ increasing until the solution either becomes singular
or the field variables $h_{k}(r), \omega _{k}(r), m(r)$ and $\sigma (r)$
have converged to constant asymptotic values to within a suitable numerical tolerance.

For soliton solutions, we start our integration close to the origin.
The need to include higher order terms in the expansions
of the gauge field functions $h_{k}(r)$ and $\omega _{k}(r)$ means that these functions
are not suitable for numerical integration.
With limited numerical precision, we cannot keep adding powers of $r$ in our initial
conditions (the expansions (\ref{eq:origin})) without losing accuracy.
For each $N$ we therefore first make a change of variables, writing the electric
gauge functions $h_{k}(r)$ in terms of new variables $\gamma _{j}(r)$, $j=1,\ldots , N-1$,
and the magnetic
gauge functions $\omega _{k}(r)$ in terms of new variables $\beta _{j}(r)$,
$j=1,\ldots , N-1$ (\ref{eq:betagamma}),
where the $\beta _{j}(r)$ and $\gamma _{j}(r)$ are chosen so that their expansions near the
origin have the form (\ref{eq:betagammaorigin}).
In Secs.~\ref{sec:su2dyons} and \ref{sec:su3dyons} the details of this change of variables
will be presented for the $N=2$ and $N=3$ cases respectively.

In the following sections, we present examples of numerical solutions of the
field equations (\ref{eq:YMefinal}--\ref{eq:Eefinal}) representing both dyons and
dyonic black holes, for $N=2$ and $N=3$.
Following \cite{Baxter:2007at}, we also study the structure of the space
of solutions by examining the phase space of parameters characterizing the solutions
near the event horizon or origin, as applicable.

In the ${\mathfrak {su}}(2)$ case, it is straightforward to show that the single electric gauge field function $h_{1}(r)$ has no zeros, as follows.
The equation for $h_{1}(r)$ takes the form (\ref{eq:hkeqn})
\begin{equation}
h_{1}''  = h_{1}' \left( \frac {\sigma '}{\sigma } - \frac {2}{r} \right)
+ \frac {2\omega _{1}^{2} h_{1}}{\mu r^{2}}.
\label{eq:h1su2}
\end{equation}
If the function $h_{1}(r)$ has a turning point at $r=r_{0}$, then $h'(r_{0})=0$
and (\ref{eq:h1su2}) gives
\begin{equation}
h_{1}''(r_{0}) = \frac {2\omega _{1}(r_{0})^{2} h_{1}(r_{0})}{\mu (r_{0}) r_{0}^{2}} .
\label{eq:h1ppsu2r0}
\end{equation}
Since the metric function $\mu (r)$ is strictly positive for all $r\ge 0$ for soliton
solutions and for all $r>r_{h}$ for black hole solutions, we conclude from
(\ref{eq:h1ppsu2r0}) that the turning point is a minimum if $h_{1}(r_{0})>0$
and a maximum if $h_{1}(r_{0})<0$.
From the expansions (\ref{eq:origin}), noting that $h_{1,0}=0$, we see that
$h_{1}'(r)$ has the same sign as $h_{1}(r)$ very close to the origin.
Similarly, near a black hole event horizon, $h_{1}'(r)$ and $h_{1}(r)$ have the same
sign from (\ref{eq:horizon}).
We deduce that it is not possible for $h_{1}(r)$ to have a turning point, and therefore
it is monotonic and has no zeros.

For ${\mathfrak {su}}(N)$, it is proven in \cite{Baxter:2015tda} that the electric gauge field functions ${\mathcal {E}}_{k}(r)$, defined in terms of the $h_{k}(r)$ by (\ref{eq:calEdef}), are monotonic and have no zeros.
While it is not necessarily the case that our alternative electric gauge field functions $h_{k}(r)$ have no zeros, since all the ${\mathcal {E}}_{k}(r)$ are nodeless any zeros of $h_{k}(r)$ are a quirk of our parametrization of the electric part of the gauge potential, rather than revealing any underlying structure of the space of solutions.
We therefore divide our numerical solutions into classes depending
on the numbers of zeros $n_{k}$ of the magnetic gauge field functions $\omega _{k}(r)$ respectively.
We use coloured plots to show this phase space structure, but hope that the
key features will still be apparent to readers using black and white.

\subsection{${\mathfrak {su}}(2)$ solutions}
\label{sec:su2}

Dyons and dyonic black holes in ${\mathfrak {su}}(2)$ EYM in adS were first found
by Bjoraker and Hosotani \cite{Bjoraker:1999yd}.
In this section we study the phase space of ${\mathfrak {su}}(2)$ solutions, checking
that we reproduce the results of \cite{Bjoraker:1999yd} and exploring
in more detail those key features which will extend to the larger $N$ case.

\subsubsection{${\mathfrak {su}}(2)$ dyonic black holes}
\label{sec:su2BH}

Four parameters are required to describe the ${\mathfrak {su}}(2)$ black hole solutions:
$r_{h}$, $\Lambda $, $h_{1}'(r_{h})$ and $\omega _{1}(r_{h})$ (\ref{eq:horizon}, \ref{eq:horexp}).
We fix $r_{h}=1$.

\begin{figure}
\begin{center}
\includegraphics[width=8.5cm]{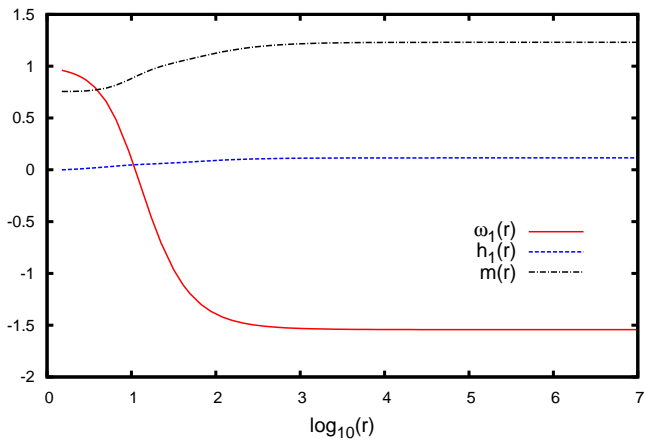}
\end{center}
\caption{Dyonic black hole solution of ${\mathfrak {su}}(2)$ EYM in adS.
The parameters are:  $r_{h}=1$, $\Lambda = -0.01$, $h_{1}'(r_{h}) = 0.01$
and $\omega _{1}(r_{h})= 0.95$.
The electric gauge field function $h_{1}(r)$ is monotonic and nodeless;
the magnetic gauge field function $\omega _{1}(r)$ has a single zero.}
\label{fig1}
\end{figure}

A typical black hole solution with $\Lambda = -0.01$ is shown in Fig.~\ref{fig1}.
As anticipated, the electric gauge field function $h_{1}(r)$ is monotonic and has
no zeros.
In Fig.~\ref{fig1}, we have chosen the initial values $h_{1}'(r_{h}) = 0.01$ and
$\omega _{1}(r_{h}) = 0.95$.
For these initial values we see that the magnetic gauge field function $\omega _{1}(r)$
has a single zero.

We now study the phase space by fixing $\Lambda $ and varying $h_{1}'(r_{h})$
and $\omega _{1}(r_{h})$.
Setting $h_{1}'(r_{h})=0$ gives purely magnetic solutions.
The field equations (\ref{eq:YMefinal}--\ref{eq:Eefinal}) are invariant
under the separate transformations $h_{1}(r) \rightarrow -h_{1}(r)$ and
$\omega _{1}(r) \rightarrow -\omega _{1}(r)$.
Therefore it suffices to consider $h_{1}'(r_{h})>0$ and $\omega _{1}(r_{h})>0$.
The values of $h_{1}'(r_{h})$ and $\omega _{1}(r_{h})$ are not completely free:
the constraint (\ref{eq:murh}) must be satisfied.
As in the purely magnetic case \cite{Baxter:2007at}, for each value of $\Lambda $
studied we find a region in the
$\left( h_{1}'(r_{h}), \omega _{1}(r_{h})\right) $-plane for which (\ref{eq:murh})
is satisfied, but we are unable to find regular black hole solutions which converge
as $r\rightarrow \infty $.

\begin{figure}
\begin{center}
\includegraphics[angle=270,width=8.5cm]{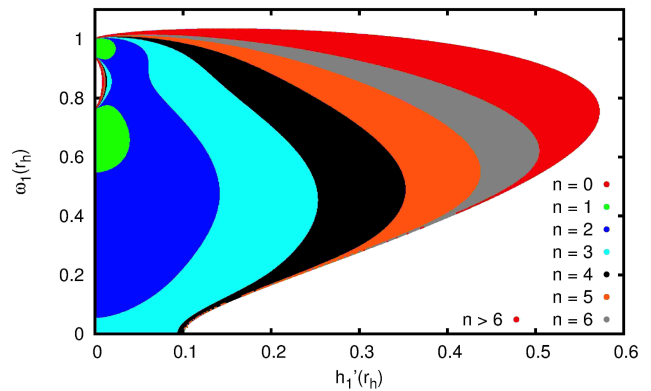}
\end{center}
\caption{Phase space of dyonic black hole solutions of ${\mathfrak {su}}(2)$ EYM,
with $r_{h}=1$ and $\Lambda = -0.01$. All shaded points in the plot correspond
to black hole solutions.  The solutions are indexed by $n=n_{1}$, the number of zeros of the
magnetic gauge field function $\omega _{1}(r)$.
The different values of $n=n_{1}$ are indicated by colour-coding the regions - in black and
white the different colours are different shades of grey.
Solutions with the largest values of $n=n_{1}$ are found towards the right-hand-side of the
coloured region.}
\label{fig2}
\end{figure}

\begin{figure}
\begin{center}
\includegraphics[angle=270,width=8.5cm]{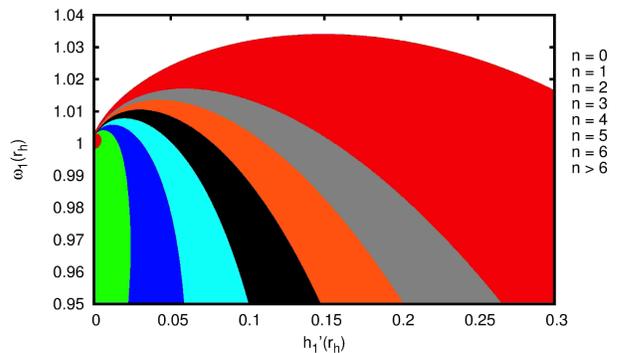}
\end{center}
\caption{Close-up view of the region surrounding the nodeless $n=n_{1}=0$ solutions
from Fig.~\ref{fig2}.
All shaded points in the plot correspond to black hole solutions.
The $n=n_{1}=0$ region (the small red or dark grey region near the point (0,1)) is in agreement with Ref.~\cite{Bjoraker:1999yd}.}
\label{fig3}
\end{figure}

In Fig.~\ref{fig2} we show the phase space for black holes with $r_{h}=1$ and
$\Lambda = -0.01$, part of which has previously been shown in \cite{Bjoraker:1999yd}.
All points in the plot in Fig.~\ref{fig2} represent black hole solutions
with particular values of $h_{1}'(r_{h})$ and $\omega _{1}(r_{h})$.
In Fig.~\ref{fig2} we find a richly structured parameter space, with solutions for which $\omega _{1}(r)$
has up to 17 nodes.
The number of zeros of $\omega _{1}(r)$ increases as $h_{1}'(r_{h})$ increases for
each fixed value of $\omega _{1}(r_{h})$.
The corresponding plot in \cite{Bjoraker:1999yd} focused on the small region near
$h_{1}'(r_{h})=0$, $\omega _{1}(r_{h})=1$ for which there are nodeless $n=n_{1}=0$ solutions,
so for comparison with \cite{Bjoraker:1999yd}, in Fig.~\ref{fig3} we show a close-up of
the parameter space near the $n=n_{1}=0$ region, which is in agreement with
\cite{Bjoraker:1999yd}.

\begin{figure}
\begin{center}
\includegraphics[angle=270,width=8.5cm]{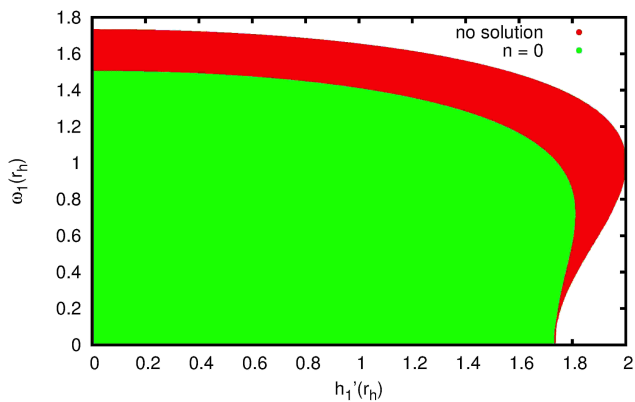}
\end{center}
\caption{Phase space of dyonic black hole solutions of ${\mathfrak {su}}(2)$ EYM,
with $r_{h}=1$ and $\Lambda = -3$.
All black hole solutions that we find have $n=n_{1}=0$.
The region labeled ``no solution'' (the red/darker grey region containing larger values of
$h_{1}'(r_{h})$ and $\omega _{1}(r_{h})$) is the region where the constraint
(\ref{eq:murh}) is satisfied, giving a regular event horizon, but we have not found
any regular black hole solutions.}
\label{fig4}
\end{figure}

For purely magnetic solutions, increasing the magnitude of the negative cosmological
constant $\left| \Lambda \right| $ increased the size of the region of
phase space where solutions were found \cite{Baxter:2007at}.
The $n=n_{1}=0$ region of nodeless solutions also expanded as a proportion of the total
solution space \cite{Baxter:2007at}.
We find the same effects for dyonic black hole solutions.
To illustrate this, in Fig.~\ref{fig4} we show the phase space of solutions
for $\Lambda = -3$ and $r_{h}=1$.
In this case the only solutions we find are such that the magnetic gauge field
function $\omega _{1}(r)$ has no zeros.
In Fig.~\ref{fig4}, we have also shown the region of the
$\left( h_{1}'(r_{h}), \omega _{1}(r_{h})\right) $-plane where the constraint
(\ref{eq:murh}) is satisfied but we have not been able to find regular solutions.
This region is marked ``no solution'' in Fig.~\ref{fig4}.

\begin{figure}
\begin{center}
\includegraphics[angle=270,width=8.5cm]{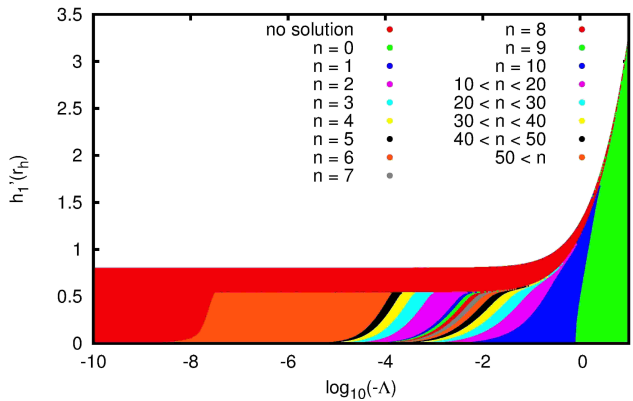}
\end{center}
\caption{Phase space of dyonic black hole solutions of ${\mathfrak {su}}(2)$ EYM
with $r_{h}=1$ and $\omega _{1}(r_{h})=0.632206952$.
The solutions are indexed by $n=n_{1}$, the number of zeros of the magnetic gauge
function $\omega _{1}(r)$.
Regions with different colours (or shades of grey) correspond to different
values of $n=n_{1}$.
We find solutions with very large values of $n=n_{1}$ as $\left| \Lambda \right| $
decreases.
The blue/darkest region (second from the right) corresponds to $n=n_{1}=1$ black hole solutions, and extends to
$\left| \Lambda \right| \rightarrow 0$, with $h_{1}'(r_{h}) \rightarrow 0$
in this limit.}
\label{fig5}
\end{figure}

\begin{figure}
\begin{center}
\includegraphics[angle=270,width=8.5cm]{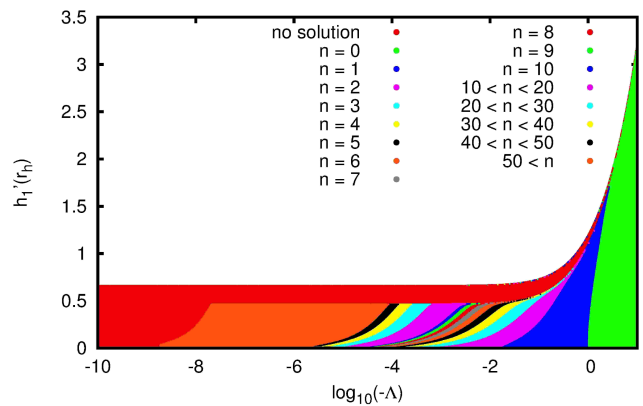}
\end{center}
\caption{Phase space of dyonic black hole solutions of ${\mathfrak {su}}(2)$ EYM
with $r_{h}=1$ and $\omega _{1}(r_{h})=0.5$.
The solutions are indexed by $n=n_{1}$, the number of zeros of the magnetic gauge
function $\omega _{1}(r)$.
Regions with different colours (or shades of grey) correspond to different
values of $n=n_{1}$.
We find solutions with very large values of $n=n_{1}$ as $\left| \Lambda \right| $
decreases.
However, the phase space of solutions does not extend to $\left| \Lambda \right| \rightarrow 0$ as there is no solution in this limit for this value of $\omega _{1}(r_{h})$.}
\label{fig6}
\end{figure}

Bjoraker and Hosotani \cite{Bjoraker:1999yd} found a rich, ``fractal''-like, structure
in the phase space of solutions as $\left| \Lambda \right| \rightarrow 0$ (see also \cite{Breitenlohner:2003qj} for similar behaviour
for solitons in the  purely magnetic case).
In asymptotically flat space, there are no dyonic black hole solutions
of ${\mathfrak {su}}(2)$ EYM \cite{Ershov:1991nv}.
However, there are purely magnetic solutions found by setting the electric part of the
gauge potential equal to zero \cite{Bizon:1990sr}.
In Figs.~\ref{fig5} and \ref{fig6} we investigate the structure of the phase space
as $\left| \Lambda \right| \rightarrow 0$.
We fix $\omega _{1}(r_{h})$ and vary $h_{1}'(r_{h})$ and $\Lambda $.
In Fig.~\ref{fig5} we set $\omega _{1}(r_{h}) = 0.632206952$, which is the value for
the first ``coloured'' black hole solution which exists in the limit
$\left| \Lambda \right| \rightarrow 0$ \cite{Bizon:1990sr}.
In Fig.~\ref{fig6} we set $\omega _{1}(r_{h})=0.5$, which does not correspond to
a regular black hole solution in the limit $\left| \Lambda \right| \rightarrow 0$.

In both Figs.~\ref{fig5} and \ref{fig6}, we have marked ``no solution'' the region where the constraint (\ref{eq:murh})
for a regular event horizon is satisfied, but our numerical solution becomes singular before $r\rightarrow \infty $.
For $\left| \Lambda \right| $ sufficiently large, for both values of $\omega _{1}(r_{h})$ the solutions are such that $\omega _{1}(r)$ is nodeless.
As $\left| \Lambda \right| $ decreases, the number of zeros of $\omega _{1}(r)$ increases rapidly.
We find solutions for which $\omega _{1}(r)$ has more than fifty zeros.
The phase spaces shown in Figs.~\ref{fig5} and \ref{fig6} are subtly different when $\left| \Lambda \right| $ is very small.
When $\omega _{1}(r_{h}) = 0.632206952$ (Fig.~\ref{fig5}), the $n=n_{1}=1$ part of the phase space (the second region from the right) extends to $\left| \Lambda \right| \rightarrow 0$ as we have the first ``coloured'' black hole solution in this limit \cite{Bizon:1990sr}.
However, for $\omega _{1}(r_{h})=0.5$, there is no solution in the limit $\left| \Lambda \right| \rightarrow 0$ and so the phase space ends at a small but nonzero value of $\left| \Lambda \right| $.
We are unable to find solutions for $\omega _{1}(r_{h})=0.5$ and $\left| \Lambda \right| $ smaller than about $10^{-9}$.

\subsubsection{${\mathfrak {su}}(2)$ dyons}
\label{sec:su2dyons}

As well as the cosmological constant $\Lambda $, dyonic solitons in ${\mathfrak {su}}(2)$ EYM are parameterized by the quantities
$\omega _{1,2}$ and $h_{1,1}$, so that the expansions (\ref{eq:origin}) of the gauge field functions take the form
\begin{eqnarray}
\omega _{1} (r) & = & 1 + \omega _{1,2}r^{2} + {\mathcal {O}}(r^{3}),
\nonumber \\
h_{1}(r) & = & h_{1,1}r + {\mathcal {O}}(r^{2}).
\end{eqnarray}
Unlike the black hole case, where the constraint (\ref{eq:murh}) for a regular nonextremal event horizon restricts the values of the parameters describing the solutions, for solitons there are no {\it {a priori}} constraints on the values that $\omega _{1,2}$ and $h_{1,1}$ can take.
In particular, $\omega _{1,2}$ can take either positive or negative values.
However, since the field equations are invariant under the transformation $h_{1}(r) \rightarrow - h_{1}(r)$, we can restrict attention to $h_{1,1}>0$ without loss of generality.

To avoid numerical errors in terms in the field equations (\ref{eq:YMefinal}--\ref{eq:Eefinal}), we define a new variable $\psi (r)$ by
\begin{equation}
\psi (r) = \omega _{1}(r)^{2} - 1,
\label{eq:psidef}
\end{equation}
which satisfies the first order differential equation
\begin{equation}
\psi '(r) = 2\omega _{1}(r) \omega _{1}'(r),
\end{equation}
and add this differential equation to those (\ref{eq:YMefinal}--\ref{eq:Eefinal}) to be integrated numerically.
Near the origin,
\begin{equation}
\psi (r) = 2\omega _{1,2}r^{2} + {\mathcal {O}}(r^{3}).
\end{equation}
In the ${\mathfrak {su}}(2)$ case, the relation (\ref{eq:betagamma}) for the gauge field functions takes the form
\begin{equation}
h_{1}(r) = \gamma _{1}(r), \qquad
\omega _{1}(r) = 1 + \beta _{1}(r) .
\end{equation}
Therefore our new variable $\psi (r)$ (\ref{eq:psidef}) is related to $\beta _{1}(r)$ by
\begin{equation}
\psi (r) = 2\beta _{1}(r) + \beta _{1}(r)^{2},
\end{equation}
and, using (\ref{eq:betagammaorigin}),
\begin{equation}
\omega _{1,2} = b_{1}.
\end{equation}

\begin{figure}
\begin{center}
\includegraphics[width=8.5cm]{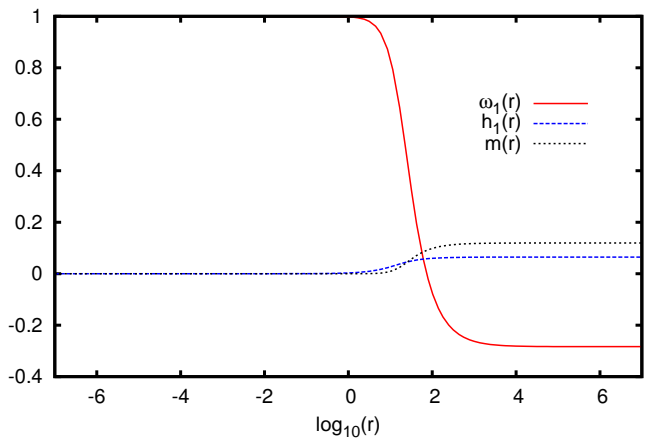}
\end{center}
\caption{Dyonic soliton solution of ${\mathfrak {su}}(2)$ EYM in adS.
The parameters are:  $\Lambda = -0.01$, $\omega _{1,2}=-0.002$, $h_{1,1}=0.003$.
The electric gauge field function $h_{1}(r)$ is monotonic and nodeless;
the magnetic gauge field function $\omega _{1}(r)$ has a single zero.}
\label{fig7}
\end{figure}

A typical soliton solution with $\Lambda = -0.01$ is shown in Fig.~\ref{fig7}, where the parameters are $\omega _{1,2}= -0.002$ and
$h_{1,1}=0.003$.
For these initial values we see that the magnetic gauge field function $\omega _{1}(r)$
has a single zero.  As expected, the electric gauge field function $h_{1}(r)$ is monotonic and has no zeros.

\begin{figure}
\begin{center}
\includegraphics[angle=270,width=8.5cm]{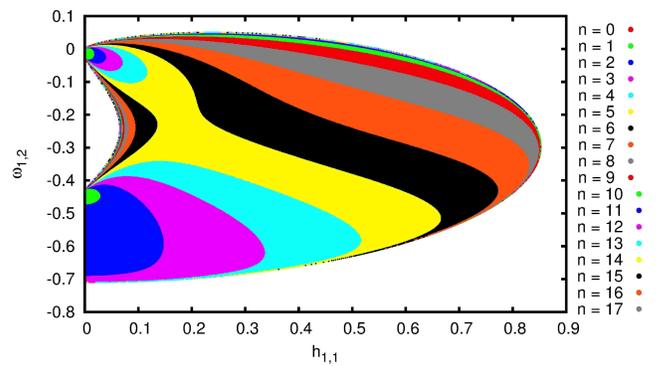}
\end{center}
\caption{Phase space of dyonic soliton solutions of ${\mathfrak {su}}(2)$ EYM,
with $\Lambda = -0.01$. All shaded points in the plot correspond
to soliton solutions.  The solutions are indexed by $n=n_{1}$, the number of zeros of the
magnetic gauge field function $\omega _{1}(r)$.
The different values of $n=n_{1}$ are indicated by colour-coding the regions - in black and
white the different colours are different shades of grey.
Solutions with the largest values of $n=n_{1}$ are found towards the right-hand-side of the
coloured region.}
\label{fig8}
\end{figure}

\begin{figure}
\begin{center}
\includegraphics[angle=270,width=8.5cm]{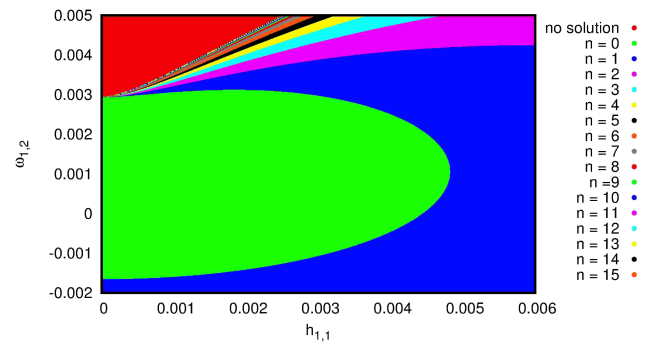}
\end{center}
\caption{Close-up view of the region surrounding the nodeless $n=n_{1}=0$ solutions
from Fig.~\ref{fig8}.
The $n=n_{1}=0$ region (the green or light grey region containing the point (0,0)) is in agreement with Ref.~\cite{Bjoraker:1999yd}.
The region labelled ``no solution'' (the red/darker grey region containing larger values of
$\omega _{1,2}$ for smaller values of $h_{1,1}$) is the region where we have not found
any regular soliton solutions.}
\label{fig9}
\end{figure}

The entire phase space of soliton solutions for $\Lambda = -0.01$ is shown in Fig.~\ref{fig8}.
All points in the plot in Fig.~\ref{fig8} represent dyonic soliton solutions
with particular values of $h_{1,1}$ and $\omega _{1,2}$.
In Fig.~\ref{fig8}, as in the black hole case (Fig.~\ref{fig2}), the phase space is very complicated, and we find solutions for which $\omega _{1}(r)$
has up to 17 nodes.
The corresponding plot in \cite{Bjoraker:1999yd} focused on the small region near
$h_{1,1}=0$, $\omega _{1,2}=0$ for which there are nodeless $n=n_{1}=0$ solutions.
For comparison, in Fig.~\ref{fig9} we show a close-up of
the parameter space near the $n=n_{1}=0$ region, which is in agreement with the corresponding plot in
\cite{Bjoraker:1999yd}.

\begin{figure}
\begin{center}
\includegraphics[angle=270,width=8.5cm]{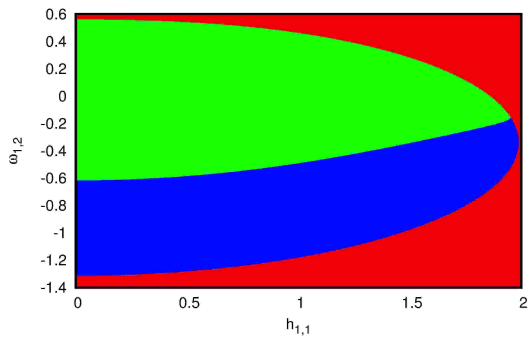}
\end{center}
\caption{Phase space of dyonic soliton solutions of ${\mathfrak {su}}(2)$ EYM,
with $\Lambda = -3$.
The different values of $n=n_{1}$ are indicated by colour-coding the regions - in black and
white the different colours are different shades of grey.
The region labelled ``no solution'' (the red/mid grey region containing larger values of
$\omega _{1,2}$ and $h_{1,1}$) is the region where we have not found
any regular soliton solutions.
There are solutions for which the number of nodes $n=n_{1}$ of the magnetic gauge field function $\omega _{1}(r)$ is either zero or one.}
\label{fig10}
\end{figure}

As $\left| \Lambda \right| $ increases, the phase space of dyonic soliton solutions simplifies, in the same way as we observed for the black hole solutions.
This can be seen in Fig.~\ref{fig10}, where we plot the phase space of  solutions for $\Lambda = -3$.
We find solutions where the magnetic gauge field function $\omega _{1}(r)$ has either zero nodes or one node.

\subsection{${\mathfrak {su}}(3)$ solutions}
\label{sec:su3}

Having discussed the phase space of ${\mathfrak {su}}(2)$ dyonic black holes and solitons in some detail, we now present new ${\mathfrak {su}}(3)$ dyonic black holes and solitons, and explore the phase space of solutions.

\subsubsection{${\mathfrak {su}}(3)$ dyonic black holes}
\label{sec:su3BH}

Dyonic black hole solutions of ${\mathfrak {su}}(3)$ EYM are described by the following six parameters: $r_{h}$, $\Lambda $, $h_{1}'(r_{h})$, $h_{2}'(r_{h})$, $\omega _{1}(r_{h})$ and $\omega _{2}(r_{h})$. We fix the event horizon radius $r_{h}=1$.
The field equations (\ref{eq:YMefinal}--\ref{eq:Eefinal}) are symmetric under the transformations $h_{k}\rightarrow -h_{k}$, $\omega _{k}\rightarrow -\omega _{k}$, for each function separately, and so we can consider $h_{k}'(r_{h})>0$, $\omega _{k}(r_{h})>0$, for $k=1, 2$ without loss of generality.

\begin{figure}
\begin{center}
\includegraphics[width=8.5cm]{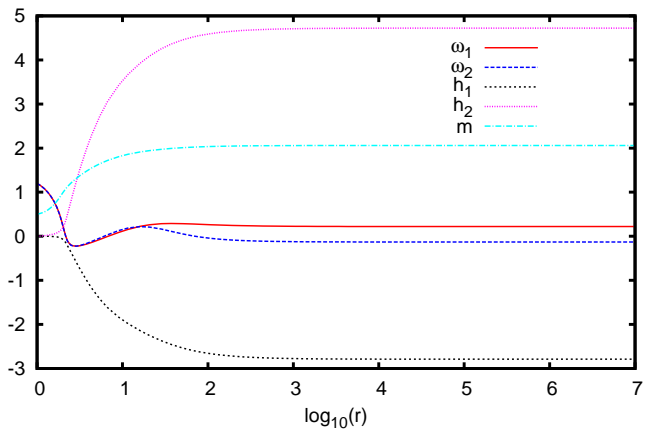}
\end{center}
\caption{Dyonic black hole solution of ${\mathfrak {su}}(3)$ EYM in adS.
The parameters are:  $r_{h}=1$, $\Lambda = -0.01$, $h_{1}'(r_{h}) = 0.01$, $h_{2}'(r_{h}) = 0.005$
and $\omega _{1}(r_{h})= 1.2 = \omega _{2}(r_{h})$.
Both electric gauge field functions $h_{1}(r)$ and $h_{2}(r)$ are monotonic and nodeless.
The magnetic gauge field function $\omega _{1}(r)$ has two zeros and $\omega _{2}(r)$ has three.}
\label{fig11}
\end{figure}

Fig.~\ref{fig11} shows a typical ${\mathfrak {su}}(3)$ dyonic black hole solution, with $\Lambda = -0.01$ and the initial values
$h_{1}'(r_{h})=0.01$, $h_{2}'(r_{h})=0.005$, $\omega _{1}(r_{h})=1.2 = \omega _{2}(r_{h})$.
For this solution the two electric gauge field functions are monotonic and have no zeros; $h_{1}(r)$ is monotonically decreasing and $h_{2}(r)$ monotonically increasing.
The two magnetic gauge field functions $\omega _{1}(r)$ and $\omega _{2}(r)$ have zeros; $\omega _{1}(r)$ has two zeros and $\omega _{2}(r)$ has three.

\begin{figure}
\begin{center}
\includegraphics[angle=270,width=8.5cm]{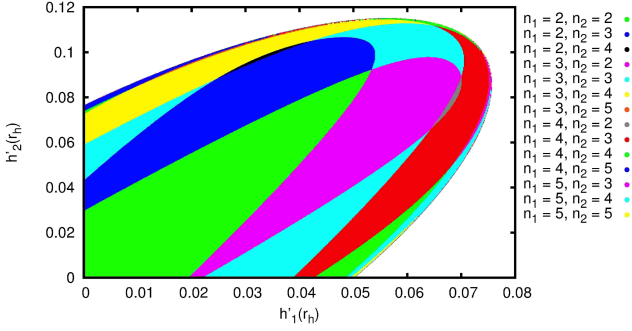}
\end{center}
\caption{Phase space of dyonic black hole solutions of ${\mathfrak {su}}(3)$ EYM with $r_{h}=1$, $\Lambda = -0.01$, and $\omega _{1}(r_{h})=1.2=\omega _{2}(r_{h})$.
All shaded points in the plot correspond to black hole solutions.
The solutions are indexed by $(n_{1}, n_{2})$, the number of zeros of the
magnetic gauge field functions $\omega _{1}(r)$, $\omega _{2}(r)$ respectively.
The different combinations of values of $(n_{1},n_{2})$ are indicated by colour-coding the regions - in black and
white the different colours are different shades of grey.
In general the number of zeros of the magnetic gauge field functions increases as we move towards the edges of the phase space.
For these values of the parameters, there are no nodeless solutions.  The number of zeros of the magnetic gauge field functions is $n_{1}=2=n_{2}$ in the green/lighter grey region containing the origin.}
\label{fig12}
\end{figure}

\begin{figure}
\begin{center}
\includegraphics[angle=270,width=8.5cm]{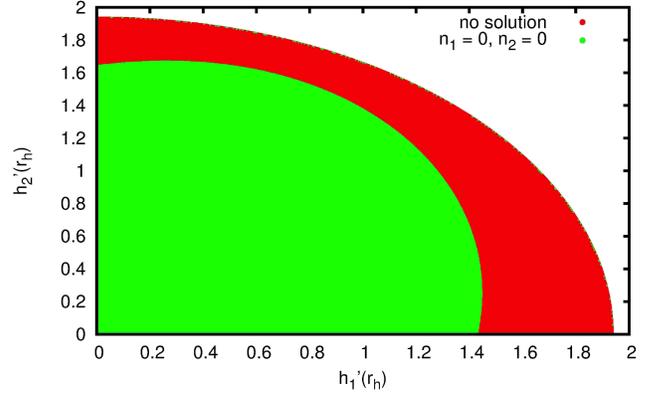}
\end{center}
\caption{Phase space of dyonic black hole solutions of ${\mathfrak {su}}(3)$ EYM with $r_{h}=1$, $\Lambda = -3$, $\omega _{1}(r_{h})=1.3$ and $\omega _{2}(r_{h})=1.2$.
The solutions are indexed by $(n_{1}, n_{2})$, the number of zeros of the
magnetic gauge field functions $\omega _{1}(r)$, $\omega _{2}(r)$ respectively.
The different combinations of values of $(n_{1},n_{2})$ are indicated by colour-coding the regions - in black and
white the different colours are different shades of grey.
For these values of the parameters, all nontrivial solutions are nodeless. We have also shown in red/darker grey (marked ``no solution'') those values of the parameters $(h_{1}'(r_{h}), h_{2}'(r_{h}))$ for which the constraint (\ref{eq:murh}) for a regular event horizon holds, but we have not found regular solutions.}
\label{fig13}
\end{figure}

In order to explore the phase space using two-dimensional plots, it is necessary to fix two of the four parameters  $h_{1}'(r_{h})$, $h_{2}'(r_{h})$, $\omega _{1}(r_{h})$ and $\omega _{2}(r_{h})$ as well as the event horizon radius $r_{h}$ and cosmological constant $\Lambda $.
Overall, the structure of the phase space of solutions is extremely complicated; we give a flavour of some of the key features in Figs.~\ref{fig12} and \ref{fig13}.
As explained in Sec.~\ref{sec:numerics}, we can classify the solutions according to the numbers of zeros $n_{1}$ and $n_{2}$ of the magnetic gauge field functions $\omega _{1}(r)$ and $\omega _{2}(r)$ respectively.
In both Figs.~\ref{fig12} and \ref{fig13}, we have fixed the values of $\omega _{1}(r_{h})$ and $\omega _{2}(r_{h})$, scanning the values of $h_{1}'(r_{h})$ and $h_{2}'(r_{h})$ such that the constraint (\ref{eq:murh}) is satisfied.
As in the ${\mathfrak {su}}(2)$ case, there are values of the parameters such that (\ref{eq:murh}) is satisfied but for which we do not find regular black hole solutions.

When the magnitude of the cosmological constant is small, we typically find a rich phase space structure with solutions with many different numbers of nodes.
This is illustrated in Fig.~\ref{fig12}, where we set $\Lambda = -0.01$.  We have also fixed the values of the magnetic gauge field functions on the event horizon to be $\omega _{1}(r_{h})=1.2=\omega _{2}(r_{h})$, and scanned the phase space by varying $h_{1}'(r_{h})$, $h_{2}'(r_{h})$.
With these values of the parameters, for all the solutions both magnetic gauge field functions $\omega _{1}(r)$ and $\omega _{2}(r)$ have at least two zeros.  As we have set the magnetic gauge field functions to be equal on the event horizon, from (\ref{eq:su2gauge}, \ref{eq:ABsu2}) there are embedded ${\mathfrak {su}}(2)$ solutions along the line $h_{2}'(r_{h}) = {\sqrt {3}}h_{1}'(r_{h})$.
In \cite{Baxter:2015tda} the existence of dyonic black hole solutions of ${\mathfrak {su}}(N)$ EYM in adS in a neighbourhood of embedded ${\mathfrak {su}}(2)$ dyonic black holes is proven.  In Fig.~\ref{fig12} we see the neighbourhood of the embedded ${\mathfrak {su}}(2)$ solutions for which there are nontrivial ${\mathfrak {su}}(3)$ dyonic black holes.
Also from Fig.~\ref{fig12}, the number of nodes of the magnetic gauge field functions increases as $h_{1}'(r_{h})$ and $h_{2}'(r_{h})$ increase.

As the magnitude of the cosmological constant increases, for ${\mathfrak {su}}(3)$ black hole solutions we find (as for the ${\mathfrak {su}}(2)$ case)  that the phase space simplifies considerably. This is illustrated in Fig.~\ref{fig13}, where we have fixed the cosmological constant to be $\Lambda = -3$ and the magnetic gauge field functions on the horizon have the values $\omega _{1}(r_{h})=1.3$, $\omega _{2}(r_{h})=1.2$.
For these values of the parameters, all nontrivial solutions are such that the two magnetic gauge field functions $\omega _{1}(r)$ and $\omega _{2}(r)$ have no zeros.
Since $\omega _{1}(r_{h})\neq \omega _{2}(r_{h})$ in Fig.~\ref{fig13}, there are no embedded ${\mathfrak {su}}(2)$ solutions in this part of the phase space.

\subsubsection{${\mathfrak {su}}(3)$ dyons}
\label{sec:su3dyons}

For ${\mathfrak {su}}(3)$ gauge group, the electric and magnetic gauge field functions need to be expanded to ${\mathcal {O}}(r^{3})$ in a neighbourhood of the origin, as follows (see the discussion in Sec.~\ref{sec:origin}):
\begin{eqnarray}
h_{k}(r) & = & h_{k,1}r+ h_{k,2}r^{2} + {\mathcal {O}}(r^{3}),
\nonumber \\
\omega _{k}(r) & = & {\sqrt {k\left( 3-k \right)}} + \omega _{k,2}r^{2} + \omega _{k,3} r^{3} + {\mathcal {O}}(r^{4}),
\nonumber \\ & &
\label{eq:su3origin}
\end{eqnarray}
for $k=1, 2$, where we have assumed without loss of generality that $\omega _{k}(0)> 0$.

The vectors ${\bmath {\omega }}_{2}= \left( \omega _{1,2}, \omega _{2,2} \right) ^{T}$ and ${\bmath {\omega }}_{3}=\left( \omega _{1,3}, \omega _{2,3} \right) ^{T}$ are eigenvectors of the matrix ${\mathcal {M}}_{2}$ (\ref{eq:calM}) with eigenvalues $2$ and $6$ respectively.
For $N=3$, the matrix ${\mathcal {M}}_{2}$ simplifies to
\begin{equation}
{\mathcal {M}}_{2} = \left(
\begin{array}{cc}
4 & - 2 \\
-2 & 4
\end{array}
\right) ,
\end{equation}
and the relevant normalized eigenvectors are \cite{Baxter:2007at}
\begin{equation}
{\bmath {u}}_{1} = \frac {1}{{\sqrt {2}}} \left(
\begin{array}{c}
1 \\ 1
\end{array}
\right) ,
\qquad
{\bmath {u}}_{2} = \frac {1}{{\sqrt {2}}} \left(
\begin{array}{c}
1 \\ -1
\end{array}
\right) ,
\label{eq:u1u2}
\end{equation}
in terms of which ${\bmath {\omega }}_{2}$ and ${\bmath {\omega }}_{2}$ are given by (\ref{eq:omega23sol}).

Similarly, the vectors ${\bmath {h}}_{1}=\left( h_{1,1}, h_{2,1} \right) ^{T}$ and ${\bmath {h}}_{2}=\left( h_{1,2}, h_{2,2} \right) ^{T}$ are eigenvectors of the matrix ${\mathcal {N}}_{2}$ (\ref{eq:calN}) with eigenvalues $2$ and $6$ respectively.
For $N=3$, the matrix ${\mathcal {N}}_{2}$ is
\begin{equation}
{\mathcal {N}}_{2} = \left(
\begin{array}{cc}
5 & -{\sqrt {3}} \\
-{\sqrt {3}} & 3
\end{array}
\right) ,
\end{equation}
with the relevant normalized eigenvectors (in terms of which ${\bmath {h}}_{2}$ and ${\bmath {h}}_{3}$ are given by (\ref{eq:h23sol})) being
\begin{equation}
{\bmath {v}}_{1} = \frac {1}{2} \left(
\begin{array}{c}
1 \\ {\sqrt {3}}
\end{array}
\right) ,
\qquad
{\bmath {v}}_{2} = \frac {1}{2} \left(
\begin{array}{c}
-{\sqrt {3}} \\ 1
\end{array}
\right) .
\label{eq:v1v2}
\end{equation}
It is straightforward to check that the eigenvectors ${\bmath {u}}_{j}$ (\ref{eq:u1u2}), ${\bmath {v}}_{j}$ (\ref{eq:v1v2}), $j=1, 2$, are related by
\begin{equation}
{\bmath {v}}_{1} = \frac {1}{{\sqrt {2}}} {\mathcal {F}}_{2}^{-1} {\bmath {u}}_{1},
\qquad
{\bmath {v}}_{2} = - {\sqrt {\frac {3}{2}}} {\mathcal {F}}_{2}^{-1} {\bmath {u}}_{2},
\end{equation}
where, for $N=3$, the matrix ${\mathcal {F}}_{2}$ (\ref{eq:calFdef}) takes the form
\begin{equation}
{\mathcal {F}}_{2} = \left(
\begin{array}{cc}
 1 &  0 \\
 -\frac {1}{2} & \frac {{\sqrt {3}}}{2}
\end{array}
\right) .
\end{equation}

In order to numerically integrate the field equations (\ref{eq:YMefinal}, \ref{eq:Eefinal}) with the initial conditions (\ref{eq:su3origin}), we seek new variables $\beta _{j}(r)$, $\gamma _{j}(r)$, $j=1, 2$, with the behaviour (\ref{eq:betagammaorigin}) near the origin.
The $\beta _{j}(r)$ depend only on the magnetic gauge field functions and the $\gamma _{j}(r)$ depend only on the electric gauge field functions.
Using the relations (\ref{eq:betagamma}) and the eigenvectors (\ref{eq:u1u2}, \ref{eq:v1v2}),
we define the $\beta _{j}(r)$ functions so that the magnetic gauge field functions $\omega _{j}(r)$ take the form
\begin{eqnarray}
\omega_{1} (r) & = & \sqrt{2} + \frac{1}{\sqrt{2}}\left[\beta_{1}(r) + \beta_{2}(r) \right],
\nonumber \\
\omega_{2}(r) & = & \sqrt{2} + \frac{1}{\sqrt{2}}\left[\beta_{1}(r) - \beta_{2}(r) \right],
\label{eq:omegasu3sol}
\end{eqnarray}
and the $\gamma _{j}(r)$ functions so that the electric gauge field functions $h_{j}(r)$ take the form
\begin{eqnarray}
h_{1}(r) & = & \frac{1}{2}\gamma_{1}(r) - \frac{\sqrt{3}}{2}\gamma_{2}(r) ,
\nonumber \\
h_{2}(r) & = & \frac{\sqrt{3}}{2}\gamma_{1}(r) + \frac{1}{2}\gamma_{2}(r).
\label{eq:hsu3sol}
\end{eqnarray}
The new variables $\beta _{1}(r)$, $\beta _{2}(r)$, $\gamma _{1}(r)$ and $\gamma _{2}(r)$ satisfy the following equations, which are derived from (\ref{eq:YMefinal}):
\begin{widetext}
\begin{subequations}
\label{eq:su3soleqns}
\begin{eqnarray}
\beta_1'' & = & -\left(\frac{\sigma'}{\sigma} + \frac{\mu'}{\mu}\right)\beta_1' + \frac{1}{4\mu r^2}(2 + \beta_1)(\beta_1^2 + 4\beta_1 + 7\beta_2^2)
           \nonumber \\ & &
           - \frac{1}{\sqrt{2}\sigma^2\mu^2}\left[ \sqrt{2}\left(\frac{9\gamma_1^2}{16} + \frac{3\gamma_2^2}{2}\right) + \frac{\beta_1}{\sqrt{2}}\left(\frac{9\gamma_1^2}{16} + \frac{3\gamma_2^2}{2}\right) -\frac{\sqrt{3}\beta_2\gamma_1\gamma_2}{\sqrt{2}}\right],
            \\
\beta_2'' & = & -\left(\frac{\sigma'}{\sigma} + \frac{\mu'}{\mu}\right)\beta_2' + \frac{1}{4\mu r^2}(7\beta_1^2 + 28\beta_2 + \beta_2^2 + 24)\beta_2 \nonumber\\
          &\, & - \frac{1}{\sqrt{2}\sigma^2\mu^2}\left[ \sqrt{6}\gamma_1 \gamma_2 + \frac{\sqrt{3}\beta_2\gamma_1 \gamma_2}{\sqrt{2}} - \frac{\beta_1}{\sqrt{2}}\left(\frac{9\gamma_1^2}{16} + \frac{3\gamma_2^2}{2}\right)\right],  \\
\gamma_1'' & = & \left( \frac{\sigma'}{\sigma} - \frac{2}{r}\right)\gamma_1' + \frac{2\gamma_1}{\mu r^2}
+ \frac{1}{\mu r^2}\left[\frac{1}{2}(\beta_1^2 + \beta_2^2) + 2(\beta_1 - \beta_2) - \beta_1\beta_2\right]\left(\frac{1}{2}\gamma_1 + \frac{\sqrt{3}}{2}\gamma_2\right) \nonumber \\
	   &\, & + \frac{1}{\mu r^2}\left[\frac{1}{2}(\beta_1^2 + \beta_2^2) + 2(\beta_1 + \beta_2) + \beta_1\beta_2\right]\left(\frac{1}{2}\gamma_1 - \frac{\sqrt{3}}{2}\gamma_2\right), \\
\gamma_2'' & = & \left( \frac{\sigma'}{\sigma} - \frac{2}{r}\right)\gamma_2' + \frac{6\gamma_2}{\mu r^2}
+ \frac{1}{\mu r^2}\left[\frac{1}{2}(\beta_1^2 + \beta_2^2) + 2(\beta_1 - \beta_2) - \beta_1\beta_2\right]\left(\frac{\sqrt{3}}{2}\gamma_1 + \frac{3}{2}\gamma_2\right) \nonumber\\
	   &\, & + \frac{1}{\mu r^2}\left[\frac{1}{2}(\beta_1^2 + \beta_2^2) + 2(\beta_1 + \beta_2) + \beta_1\beta_2\right]\left(\frac{\sqrt{3}}{2}\gamma_1 - \frac{3}{2}\gamma_2\right).
\end{eqnarray}
\end{subequations}
\end{widetext}
We also substitute for $h_{j}(r)$ and $\omega _{j}(r)$ from (\ref{eq:omegasu3sol}, \ref{eq:hsu3sol}) into the Einstein equations (\ref{eq:Eefinal}), and then numerically integrate the resulting equations, together with (\ref{eq:su3soleqns}), using the initial conditions (\ref{eq:betagammaorigin}).
The solutions are parametrized by the cosmological constant $\Lambda $, and the four parameters $b_{1}$, $b_{2}$, $g_{1}$ and $g_{2}$.
As in the ${\mathfrak {su}}(2)$ case, there are no {\it {a priori}} constraints on the values of these four parameters.  In general $b_{1}$ and $b_{2}$ can take both positive and negative values.
In our phase space plots, we have restricted our attention to $g_{1}>0$, $g_{2}>0$ since this reveals the key features of the phase space.

\begin{figure}
\begin{center}
\includegraphics[width=8.5cm]{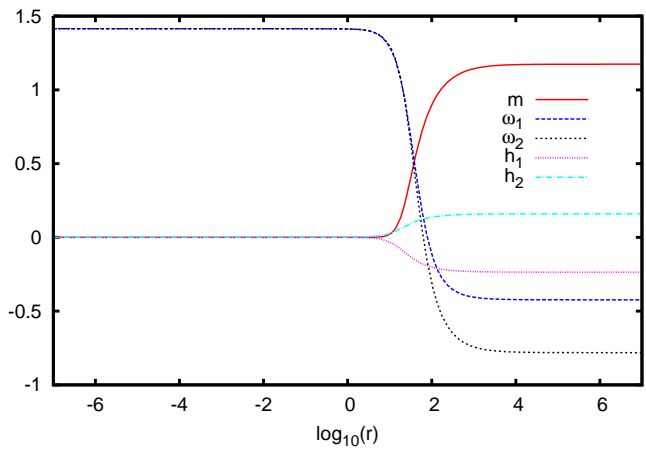}
\end{center}
\caption{Dyonic soliton solution of ${\mathfrak {su}}(3)$ EYM in adS.
The parameters are:  $\Lambda = -0.01$, $b_{1}=-0.002$, $b_{2}=-0.00001$, $g_{1}=0.001$, $g_{2}=0.0005$.
Both electric gauge field functions $h_{1}(r)$ and $h_{2}(r)$ are monotonic and nodeless.
The magnetic gauge field functions $\omega _{1}(r)$ and $\omega _{2}(r)$ both have a single zero.}
\label{fig14}
\end{figure}

A typical dyonic soliton solution of ${\mathfrak {su}}(3)$ EYM is shown in Fig.~\ref{fig14}.
The cosmological constant is $\Lambda = -0.01$ and the other parameters are $b_{1}=-0.002$, $b_{2}=-0.00001$, $g_{1}=0.001$, $g_{2}=0.0005$.
For these values of the parameters, the two electric gauge field functions $h_{1}(r)$ and $h_{2}(r)$ are monotonic and have no zeros; $h_{1}(r)$ is monotonically decreasing and $h_{2}(r)$ is monotonically increasing.
The two magnetic gauge field functions $\omega _{1}(r)$ and $\omega _{2}(r)$ are monotonically decreasing, and both have a single zero.

\begin{figure}
\begin{center}
\includegraphics[angle=270,width=8.5cm]{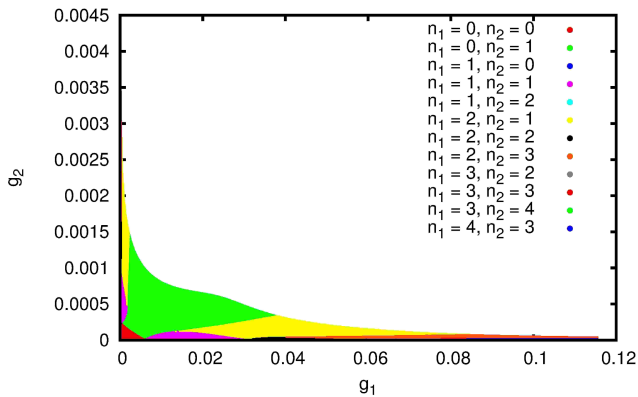}
\end{center}
\caption{Phase space of dyonic soliton solutions of ${\mathfrak {su}}(3)$ EYM with $\Lambda = -0.01$, $b_{1}=-0.002$, $b_{2}=-0.00001$.
All shaded points correspond to soliton solutions.
The solutions are indexed by $(n_{1}, n_{2})$, the number of zeros of the
magnetic gauge field functions $\omega _{1}(r)$, $\omega _{2}(r)$ respectively.
The different combinations of values of $(n_{1},n_{2})$ are indicated by colour-coding the regions - in black and
white the different colours are different shades of grey.
We find solutions where the magnetic gauge field functions have a wide variety of numbers of nodes.  There is a small region close to the origin where both
$\omega _{1}(r)$ and $\omega _{2}(r)$ have no zeros.}
\label{fig15}
\end{figure}

\begin{figure}
\begin{center}
\includegraphics[angle=270,width=8.5cm]{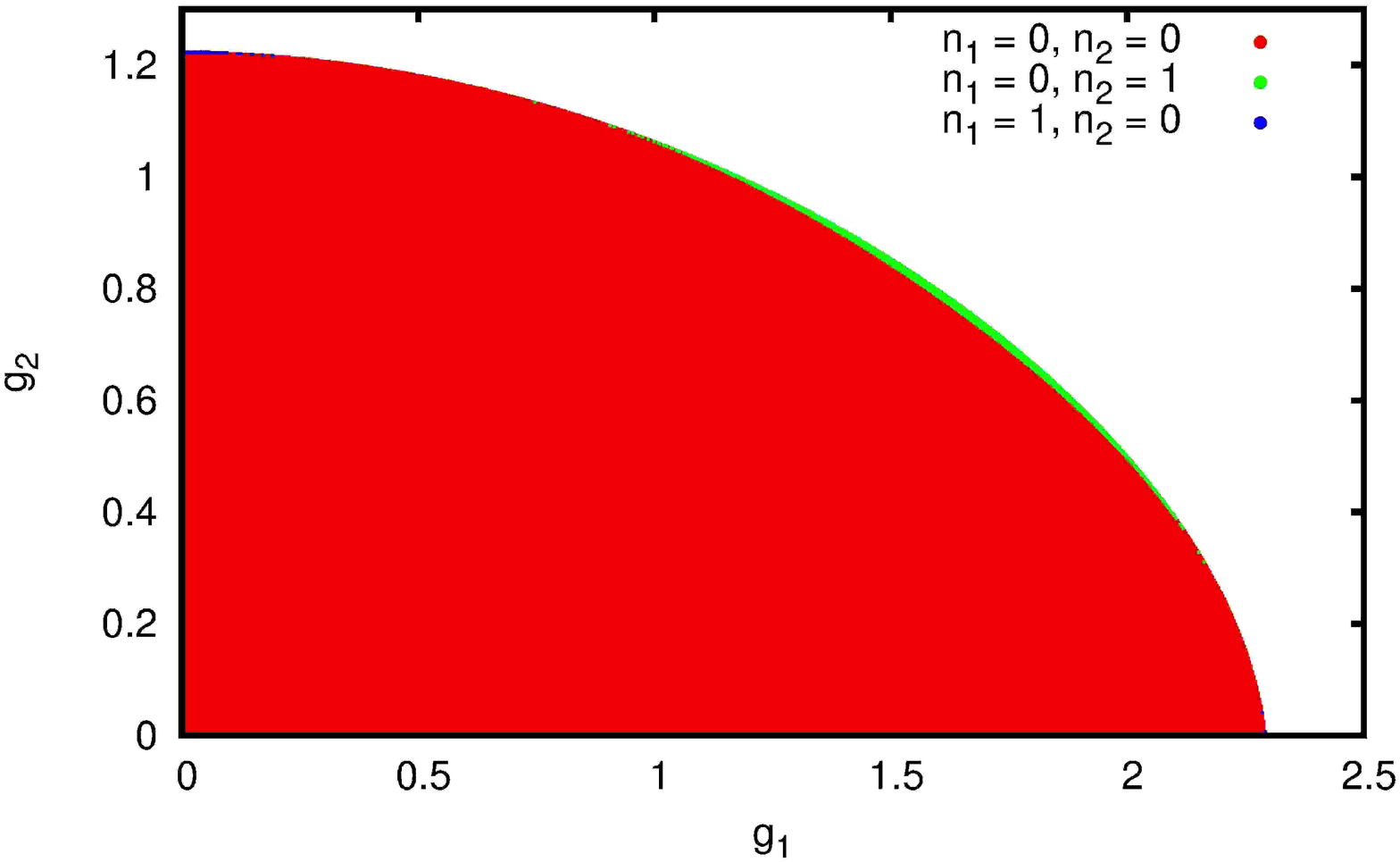}
\end{center}
\caption{Phase space of dyonic soliton solutions of ${\mathfrak {su}}(3)$ EYM with $\Lambda = -3$, $b_{1}=-0.2$, $b_{2}=-0.1$.
All shaded points correspond to soliton solutions.
The solutions are indexed by $(n_{1}, n_{2})$, the number of zeros of the
magnetic gauge field functions $\omega _{1}(r)$, $\omega _{2}(r)$ respectively.
The different combinations of values of $(n_{1},n_{2})$ are indicated by colour-coding the regions - in black and
white the different colours are different shades of grey.
The phase space has a much simpler structure for this larger value of $\left| \Lambda \right|$. For most of the solution space, both magnetic gauge field functions $\omega _{1}(r)$ and $\omega _{2}(r)$ have no zeros. There are small regions at the edges of the space of solutions where either $\omega _{1}(r)$ or $\omega _{2}(r)$ has a single zero.}
\label{fig16}
\end{figure}

Once again we find a very rich space of solutions, and illustrate some features in Figs.~\ref{fig15} and \ref{fig16}.
Fig.~\ref{fig15} shows the phase space for $\Lambda = -0.01$. The parameters $b_{1}=-0.002$ and $b_{2}=-0.00001$ which govern the behaviour of the magnetic field functions near the origin are fixed; we have scanned over positive values of the parameters $g_{1}$ and $g_{2}$ describing the electric gauge field functions near the origin.
As we have seen previously for both ${\mathfrak {su}}(2)$ solutions and ${\mathfrak {su}}(3)$ dyonic black holes, for ${\mathfrak {su}}(3)$ dyonic solitons with $\left| \Lambda \right| $ comparatively small, the phase space is very complicated. There are many different regions in which the two magnetic gauge field functions $\omega _{1}(r)$ and $\omega _{2}(r)$ have different numbers of zeros.  For these values of the parameters, we find that $\omega _{1}(r)$ and $\omega _{2}(r)$ can have up to four zeros.  We find a small region close to $g_{1}=0=g_{2}$ for which both magnetic gauge field functions have no zeros.
Embedded ${\mathfrak {su}}(2)$ solutions correspond to $b_{2}=0=g_{2}$, and therefore there are no embedded solutions in Fig.~\ref{fig15}.

As the magnitude of the cosmological constant increases, the phase space simplifies considerably. This can be seen in Fig.~\ref{fig16}, where $\Lambda = -3$.
The size of the space of nontrivial solutions also expands as $\left| \Lambda \right| $ increases.
In Fig.~\ref{fig16}, we have fixed $b_{1}=-0.2$ and $b_{2}=-0.1$, and varied $g_{1}>0$ and $g_{2}>0$.
With these values of the parameters, most of the nontrivial dyonic soliton solutions are such that $n_{1}=0=n_{2}$ and both magnetic gauge field functions have no zeros.
There are small regions close to the edge of the solution space where one of $(\omega _{1}(r), \omega _{2}(r))$ (but not both) has a single zero.

\section{Conclusions}
\label{sec:conc}

In this paper we have presented new dyonic soliton and black hole solutions of the ${\mathfrak {su}}(N)$ Einstein-Yang-Mills (EYM) field equations in asymptotically anti-de Sitter (adS) space-time with a negative cosmological constant $\Lambda <0$. The metric is static and spherically symmetric. The gauge field has nontrivial electric and magnetic components, and is described by $2(N-1)$ independent gauge field functions, with equal numbers of electric and magnetic gauge field functions.
We have explored the phase space of soliton and black hole solutions for ${\mathfrak {su}}(2)$ and ${\mathfrak {su}}(3)$ gauge groups.
The solutions can be categorized by the numbers of zeros, $n_{j}$, of the magnetic gauge field functions $\omega _{j}(r)$.
In general the phase space is very rich, with many different combinations of $n_{j}$ possible.  However, we find the following general features, many of which are in common with the phase space of purely magnetic ${\mathfrak {su}}(N)$ solutions \cite{Baxter:2007at}:
\begin{itemize}
\item
For small $\left| \Lambda \right| $, we find solutions in which the magnetic gauge field functions have large numbers of zeros, particularly for the ${\mathfrak {su}}(2)$ case;
\item
For small $\left| \Lambda \right| $, the phase space is particularly complicated, with many different combinations of values of $n_{j}$;
\item
As $\left| \Lambda \right| $ increases, the phase space expands in parameter space and the number of different combinations of values of $n_{j}$ decreases;
\item
For large $\left| \Lambda \right| $, there are solutions for which all the magnetic gauge field functions have no zeros.
\end{itemize}
The existence of nontrivial dyonic solutions has been proven recently in the following regimes in parameter space \cite{Baxter:2015tda}:
\begin{itemize}
\item
In a neighbourbood of the embedded trivial solution, either pure adS (for solitons) or Schwarzschild-adS (for black holes);
\item
In a neighbourhood of embedded nontrivial ${\mathfrak {su}}(2)$ dyonic soliton and black hole solutions;
\item
In a neighbourhood of nontrivial purely magnetic ${\mathfrak {su}}(N)$ solutions (whose existence is proven in \cite{Baxter:2008pi}).
\end{itemize}

Of particular interest are the solutions in the intersection of the last items in the two lists above, namely nontrivial nodeless solutions in the intersection of a neighbourbood of embedded ${\mathfrak {su}}(2)$ solutions and a neighbourhood of embedded purely magnetic ${\mathfrak {su}}(N)$ solutions.
We conjecture that it may be possible to prove that such solutions are stable under linear, spherically symmetric perturbations.
Recently the existence of stable dyonic soliton and black hole solutions of ${\mathfrak {su}}(2)$ EYM in adS has been proven \cite{Nolan:2015vca}, and it would be interesting to attempt to extend that analysis to the case of a larger gauge group. In the ${\mathfrak {su}}(2)$ case, the perturbation equations for dyonic solutions are much more complicated than the corresponding equations for purely magnetic background solutions, and the same will be true for ${\mathfrak {su}}(N)$ gauge group.
We therefore leave this question open for future work.

The dyonic soliton and black hole solutions studied in this paper are spherically symmetric, with the event horizon being a surface of constant positive curvature.
The existence proof in \cite{Baxter:2015tda} is more general, and applies also to topological black holes for which the event horizon has either zero or constant negative curvature.
A natural question would be to investigate the phase space of dyonic topological black hole solutions of the ${\mathfrak {su}}(N)$ EYM equations, extending the recent study of the phase space of purely magnetic topological black hole solutions \cite{Baxter:2015ffm}.
Black holes with a flat event horizon in particular have attracted a great deal of attention in the literature as models of holographic superconductors (see \cite{Cai:2015cya} for a recent review).
Planar black holes with a ${\mathfrak {su}}(2)$ gauge field have been used to model $p$-wave superconductors (see, for example, \cite{Gubser:2008wv} for a selection of papers), and enlarging the gauge group in these models would also be of interest.
We plan to return to this topic in a future publication.

Finally, we anticipate that the thermodynamics of the dyonic black holes presented here would be very interesting.
The thermodynamics of purely magnetic ${\mathfrak {su}}(2)$ black holes in adS has recently been studied \cite{Kichakova:2015lza,Fan:2014ixa}.
In \cite{Kichakova:2015lza}, a complex picture emerges: it is found that purely magnetic, spherically symmetric ${\mathfrak {su}}(2)$ black holes with unit magnetic charge at infinity are globally thermodynamically unstable; those with zero magnetic charge at infinity have two branches of solutions, both of which are globally thermodynamically unstable to decay to an embedded Schwarzschild-adS black hole; while those with general magnetic charge at infinity also have two branches of solutions, one of which is thermodynamically stable.
We expect that the additional complexity of both enlarging the gauge group and including an electric as well as a magnetic part of the gauge field will render the thermodynamics of the ${\mathfrak {su}}(N)$ dyonic black holes studied in this paper to be even more complicated than that presented in \cite{Kichakova:2015lza} for the purely magnetic ${\mathfrak {su}}(2)$ case.
Accordingly we leave a systematic study of the thermodynamics to future research.

\begin{acknowledgments}
We thank Erik Baxter for insightful discussions.
The work of B.L.S. is supported by UK EPSRC.
The work of E.W. is supported by the Lancaster-Manchester-Sheffield
Consortium for Fundamental Physics under STFC Grant No.~ST/L000520/1.
\end{acknowledgments}

\end{document}